\documentclass[a4paper,11pt]{article}
\pdfoutput=1 
\usepackage{verbatim}

\usepackage{jheppub}
\usepackage{amsmath,amsfonts,amssymb,amsthm,graphicx,subfigure,color}
\usepackage[T1]{fontenc} 
\usepackage{enumerate}

\usepackage[all]{xy}
\usepackage{amssymb}
\usepackage{amsmath}
\newcommand{\cs}{\text{cos}\,2\beta}
\newcommand{\sn}{\text{sin}\,2\beta}
\newcommand{\tn}{\text{tan}\,2\beta}
\newcommand{\scc}{\text{sec}\,2\beta}
\newcommand{\ct}{\text{cot}\,2\beta}
\newcommand{\bj}{\hat J}

\newcommand{\bom}{\hat \Omega}

\newcommand{\hr}{\hat\rho}
\newcommand{\hp}{\hat\psi}

\newcommand{\jc}{J_\mathcal{C}}
\newcommand{\pc}{P_\mathcal{C}}
\newcommand{\dc}{d_\mathcal{C}}

\title{\boldmath 	Gravity duals of $\mathcal{N}=(0,2)$ SCFTs from M5-branes}

\author[]{Ibrahima Bah, Vasilis Stylianou}
\affiliation{Department of Physics and Astronomy, University of Southern California, Los Angeles, CA 90089, USA}

\emailAdd{bah@usc.edu, styliano@usc.edu}

\abstract{We describe the general BPS system that governs the gravity duals of $\mathcal{N}=(0,2)$ two-dimensional superconformal field theories in the low-energy limit of M5-branes on a four-manifold, $M_4$.  In order to preserve supersymmetry, we restrict to cases where the four-manifold is embedded in a Calabi-Yau fourfold that is a sum of two line bundles over $M_4$.  We further reduce the $\mathcal{N}=(0,2)$ system to describe the gravity duals of SCFTs with $\mathcal{N}=(0,4)$ and $\mathcal{N}=(2,2)$ supersymmetry.  In the first case, the solutions fit in the larger class of $AdS_3 \times S^2 \times CY_3$ solutions of M-theory dual to $\mathcal{N}=(0,4)$ SCFTs.  In the case of the $\mathcal{N}=(2,2)$ theories, the near-horizon limit of $M_4$ is necessarily a product of two constant curvature Riemann surfaces whose metrics are governed by a pair of Liouville equations.     
}
\preprint{}
\date{}

\begin{document}
\maketitle
\flushbottom

%%%%%%%%%%%%%%%%%%%%%%%%%%%%%%%%
\section{Introduction}
%%%%%%%%%%%%%%%%%%%%%%%%%%%%%%%%%%%%

M5-branes in M-theory have emerged as one of the most powerful tools for studying and classifying superconformal field theories (SCFTs)  in lower dimensions.  These SCFTs arise in the low-energy dynamics of twisted compactifications of M5-branes on manifolds of various dimensions.  The possible choices of manifolds that can be wrapped by M5-branes are restricted by supersymmetry. In turn, when these geometries are understood, the protected sectors of the SCFTs can be studied in terms of the topological and geometric properties of these manifolds.  This then provides a new way to identify, study and classify SCFTs without the requirement of a Lagrangian description.  Moreover, the manifolds can also have boundaries; in which case, supersymmetric boundary conditions are needed for the M5-branes.  In the field theories, this data manifests itself as various types of global symmetries; therefore the addition of boundaries enriches the geometric classification scheme.  

In the present paper, we are interested in two-dimensional $\mathcal{N}=(0,2)$ SCFTs that arise in the low-energy dynamics of M5-branes wrapped on four-manifolds.  We study this problem in AdS/CFT  and derive the BPS system which gives rise to the holographic duals of these field theories. In particular, we start with the supergravity equations of M-theory for $AdS_3$ vacua \cite{Gauntlett:2006qw} and further reduce to cases where the vacua emerge from the near-horizon limit of a stack of M5-branes wrapped on a four-manifold, $\mathcal{C}$.  We assume that this four-manifold is generic and may admit boundaries. Furthermore, in order to preserve supersymmetry, we consider cases where the four-manifold is embedded in a Calabi-Yau fourfold ($CY_4$). We further restrict to cases where we take $CY_4$ to be a sum of two line bundles over $\mathcal{C}$.  The reduction of the supergravity equations then allows for the classification of the possible choices of $\mathcal{C}$ and boundary conditions of the M5-branes, and therefore a classification for these $\mathcal{N}=(0,2)$ SCFTs.

This program has been invigorated with the classification of four-dimensional $\mathcal{N}=2$ SCFTs in \cite{Gaiotto:2009we,Gaiotto:2009hg} by considering twisted compactifications of M5-branes on two-dimensional Riemann surfaces with punctures.   Strong evidence for the existence of the $\mathcal{N}=2$ SCFTs was provided in \cite{Gaiotto:2009gz} by explicit constructions of families of $AdS_5$ solutions dual to the $\mathcal{N}=2$ SCFTs by using the Lin, Lunin and Maldacena (LLM) system in \cite{Lin:2004nb}.  In these solutions, punctures manifest themselves as localized sources on the Riemann surface that are extended along the normal directions.  The metric on this space is governed by a single potential that satisfies a partial differential equation, the $SU(\infty)$ Toda equation; and the set of sources to the Toda equation that lead to regular solutions corresponds to the set of supersymmetric boundary conditions for the M5-branes at the punctures.  For cases where there are no punctures, the Toda equation reduces to Liouville equation, and the Riemann surface has constant curvature and decouples from the normal directions.  This program has further been extended to four-dimensional $\mathcal{N}=1$ SCFTs. The $\mathcal{N}=1$ generalizations of LLM are obtained in \cite{Bah:2013qya,Bah:2015fwa}, while the structure of general $\mathcal{N}=1$ punctures from the $AdS_5$ system is discussed in \cite{Bah:2015fwa}.

The results in the present paper can therefore be seen as generalizations of the LLM system to $AdS_3$ solutions dual to $\mathcal{N}=(0,2)$ theories from M5-branes on four-manifolds.  In particular, the equations that govern the metric need to be general enough to allow for the description of boundaries similar to the $SU(\infty)$ Toda equation.  

A class of $AdS_3$ solutions from M5-branes have been studied in \cite{Gauntlett:2000ng,Benini:2013cda}.  In both cases, the authors consider twisted compactifications of M5-branes\footnote{The tools for twisted compactifications of branes in string theory and M-theory were developed in \cite{Maldacena:2000mw} in the context of gauged supergravity.}  on a product of two Riemann surfaces with no punctures and four-manifolds with no boundaries.  Furthermore, their M-theory constructions arise as $AdS_3$ vacua in seven-dimensional gauged supergravity which are then uplifted to M-theory.  This procedure, while powerful in generating new solutions, is very restrictive in capturing systems which involve boundaries or orbifolds.  A strong motivation for this work is to capture the more complex systems that involve boundaries.

This classification question for two-dimensional SCFTs has also been studied in the context of quantum field theory and four-manifolds in \cite{Benini:2013cda,Gadde:2013sca}.  Moreover, $AdS_3$ solutions have also been obtained by using non-Abelian T-duality in \cite{Bea:2015nx,Lozano:2015qy} and gauged supergravity in lower dimensions \cite{Benini:2013cda,Cucu:2003yk,Donos:2014rt,Bobev:2014zr,Naka:ul,Karndumri:2013uq}.  Gravity duals to $\mathcal{N}=(0,4)$ SCFTs from M5-branes are also discussed in \cite{Gauntlett:2006fk,Kim:2012ek}.

 The plan of the rest of the paper is as follows. In section \ref{M5branes} we discuss generic properties of the two-dimensional $\mathcal{N}=(0,2)$ SCFTs by considering twisted compactifications of the world volume theory of a stack of $N$ M5-branes -- the six-dimensional $(2,0)$ $A_{N-1}$ SCFT. We also consider some special twists where the supersymmetry enhances to $(0,4)$ and $(2,2)$. 

In section \ref{Gravduals} we start by reviewing the geometric set-up for the M5-branes on four-manifolds in M-theory.  Then, in section \ref{02fromM5} we describe the general metric for $AdS_3$ solutions dual to $\mathcal{N}=(0,2)$ SCFTs from systems of M5-branes.  In sections \ref{ansatz} and \ref{ansatzflux} we motivate the ansatz for systems where a stack of M5-branes wrap a four-manifold, $\mathcal{C}$, which is embedded in a CY4 that is a sum of two line bundles over $\mathcal{C}$.  We end section \ref{Gravduals} with a presentation of the gravity duals in \ref{02system}.  In section \ref{enhancesym} we further reduce to systems which preserve $(0,4)$ and $(2,2)$. We conclude with a summary of our results and a discussion about the next steps in this program in section \ref{conclusion}.  The work and results in this paper are fairly technical; we do our best to hide this in the appendices.  

%%%%%%%%%%%%%%%%%%%%%%%%%%%%%%%%%%%%%%%%%%
\section{ $\mathcal{N}=(0,2)$ from M5-branes} \label{M5branes}
%%%%%%%%%%%%%%%%%%%%%%%%%%%%%%%%%%%%%%%%%%%

We start by reviewing some general properties of two-dimensional $\mathcal{N}=(0,2)$ theories obtained by compactifying a stack of $N$ $M5$-branes on a four-manifold, $M_4$.  The surface is given by a co-dimension two complex curve, $\mathcal{C}$, in a Calabi-Yau fourfold.  Equivalently, we consider the compactifications of the six-dimensional $A_{N-1}$ $(0,2)$ SCFT to two dimensions.  In particular, we will study cases where the two-dimensional theory preserves at least a $U(1)^2$ global symmetry, one of which is the $\mathcal{N}=(0,2)$ $U(1)$ R-symmetry.

%%%%%%%%%%%%%%%%%%%%%%%%%%%%%%%%%%%%%%%%%%%%%
\subsection{Twist and Symmetries}\label{twist}
%%%%%%%%%%%%%%%%%%%%%%%%%%%%%%%%%%%%%%%%%%%%%%

Supersymmetry is generically broken when supersymmetric field theories are taken over curved manifolds.  However, some supersymmetry can be preserved  by a partial topological twist \cite{Witten:1988ze,Bershadsky:1995qy}.  A constant global spinor is obtained by gauging the R-symmetry in a way that trivializes the killing spinor equation.  In other words we pick a background gauge field, $A$ valued in the R-symmetry, to solve the killing spinor equation
\begin{equation}
(\partial_{\mu}+i\omega_{\mu}-iA_{\mu})\epsilon=0
\end{equation} where $\omega^{\mu}$ is the spin connection of the manifold and $\varepsilon$ is the desired spinor.  The possible choices of twists are determined by the holonomy group of $M_4$, Hol($\nabla$), and the different ways we can embed it in the R-symmetry.

The four-manifold $M_4$ is complex and generically admits a $U(2)$ holonomy group; in this paper, we merely consider twists of its Abelian subgroup, $U(1)_1 \times U(1)_2$.  Now, the six-dimensional (2,0) theory has an $SO(5)$ R-symmetry whose Cartan subgroup is given as
\begin{equation}
U(1)_{+}\times U(1)_{-}\subset SU(2)_{+}\times SU(2)_{-}\subset SO(5).
\end{equation} Thus, the different Abelian twists, which we are interested in, are obtained by identifying the $U(1)$'s in Hol$(\nabla)$ with different combinations of $U(1)_\pm$.  More formally, we can relate the generators as
\begin{equation}
J_{\sigma}=\frac{p_{\sigma}}{p_{\sigma}+q_{\sigma}} J_{+}+\frac{q_{\sigma}}{p_{\sigma}+q_{\sigma}} J_{-},\qquad\qquad \sigma=1,2,
\end{equation}  where $J_\sigma$ and $J_\pm$ are generators of $U(1)_\sigma$ and $U(1)_\pm$ respectively.  The $(p_\sigma, q_\sigma)$ are the flux of the background gauge fields; combinations of them are fixed by the topological twist.

In the special cases, for example, where the four-manifold is a trivial product of two Riemann surfaces, $M_4 = \Sigma_1 \times \Sigma_2$,  the $U(1)_1 \times U(1)_2$ can be identified with the $U(1)$ holonomies of the surfaces respectively.   The fluxes are then fixed as
\begin{equation}\label{twistsigma}
p_{\sigma}+q_{\sigma}=- \chi\left(\Sigma_\sigma \right)
\end{equation} where $\chi\left(\Sigma_\sigma \right)$ is the Euler characteristic of the Riemann surface.  

With the twist, the bosonic symmetries of the six-dimensional theory are broken down as
\begin{equation}
SO(1,5) \times SO(5) \; \to \; SO(1,1) \times U(1)_+ \times U(1)_-
\end{equation}  and the system generically preserves two supercharges.  The field theory admits an R-symmetry, $R_0$, and  a flavor symmetry $\mathcal{F}$; these can be written in terms of the generator of $U(1)_\pm$  as
\begin{equation}
R_0 = \frac{1}{2} \left(J^+ + J^-\right), \qquad \mathcal{F} = \frac{1}{2} \left(J^+ -J^-\right).  \label{U(1)syms}
\end{equation} When the two-dimensional field theory is superconformal, we can write the superconformal R-symmetry, $R_{\mathcal{N}=(0,2)}$, as 
\begin{equation}
R_{\mathcal{N}=(0,2)} = R_0 + \epsilon \mathcal{F}= a_+ J^+ + a_- J^-, \qquad a_\pm = \frac{1}{2} \left(1\pm \epsilon \right).  \label{Rsymfield}
\end{equation}  The $\epsilon$ parameter is fixed by c-extremization when the two-dimensional SCFT is compact \cite{Benini:2012cz}.  

%%%%%%%%%%%%%%%%%%%%%%%%%%%%%%%%%%%
\subsubsection*{Enhanced symmetry}
%%%%%%%%%%%%%%%%%%%%%%%%%%%%%%%%%%%%%

For special values of the parameters, the two-dimensional theories can have enhanced symmetries.  In particular we will be interested in cases where the supersymmetry enhances to $\mathcal{N}=(0,4)$ and to $\mathcal{N}=(2,2)$ (see appendix E of \cite{Benini:2013cda} for a complete discussion of the supersymmetries for twisted compactifications of M5-branes on four-manifolds).  

\paragraph{$(0,4)$ theories} We can fix $p_1 = p_2=0$ or $q_1 =q_2=0$ to obtain $(4,0)$ or $(0,4)$ theories.  Consider the case when $p_1 =p_2=0$.  The $U(1)_+$ is not twisted and therefore it enhances to $SU(2)_+$, the R-symmetry of the $(0,4)$ theories.  The topological twist is along $U(1)_-$.  The bosonic symmetries of the six-dimensional theory is broken down as
\begin{equation}
SO(1,5) \times SO(5) \; \to \; SO(1,1) \times U(1)_- \times SU(2)_+.
\end{equation}

\paragraph{$(2,2)$ theories}  When $p_1 =q_2=0$ or $p_2 =q_1=0$, we obtain $(2,2)$ theories.  In this case, the symmetries are $U(1)_+ \times U(1)_-$ and they can be identified with the left/right R-symmetry of $(2,2)$ theories.  In these cases, only the supersymmetry enhances, the global symmetry coming from the $SO(5)$ R-symmetry does not change.

\paragraph{$(0,2)$ with $SU(2)_{\mathcal{F}} $}  When $p_\sigma =q_\sigma$, the diagonal $U(1)$ of $U_+(1) \times U_-(1)$ is not twisted, it enhances to a global $SU(2)_{\mathcal{F}} $ flavor symmetry.  This symmetry is also the diagonal $SU(2)$ of $SU(2)_+ \times SU(2)_- \subset SO(5)$.

%%%%%%%%%%%%%%%%%%%%%%%%%%%%%%%%%%%%%%%%%%%%&&&&&&&&&&&&&&
\section{Gravity Duals to $\mathcal{N}=(0,2)$ from M5-branes}\label{Gravduals}
%%%%%%%%%%%%%%%%%%%%%%%%%%%%%%%%%%%%%%%%%%%%%%%%%%%%%%%%%%

Our primary interest is to understand when twisted compactifications of M5-branes flow to two-dimensional $\mathcal{N}=(0,2)$ SCFTs in the infrared (IR) limit. In the large $N$ limit, this question can be studied in $AdS$/CFT by classifying the set of $AdS_{3}$ solutions that can exist in the near-horizon limit of systems of $N$ M5-branes wrapping a four-manifold, $M_4$. However, the problem of obtaining the full M-theory solution is in general hard, and since we are only interested in cases where the near-horizon geometry contains a decoupled $AdS_{3}$ factor, it suffices to look for these geometries directly. In this section, we discuss how to construct these $AdS_3$ solutions at a qualitative level. We refer the reader to the appendices for a more quantitative discussion.

Our starting point is M-theory where we decompose the 11d spacetime as
\begin{equation}
M^{1,10} \cong \mathbb{R}^{1,1} \times \mathbb{R} \times CY_4, \label{UVspacetime}
\end{equation}  and consider a stack of $N$ M5-branes in this background with a world-volume  $\mathbb{R}^{1,1} \times \mathcal{C}$. Here we assume that the four-manifold $M_4$ can be described by a complex co-dimension two surface, $\mathcal{C}$, in a $CY_4$. We also assume that $\mathcal{C}$ is generic and may admit boundaries.  Now, in the region near the branes, the $CY_4$ is given by a $\mathbb{C}^2$ bundle over the complex surface $\mathcal{C}$.  

In the case of interest, the structure group of the bundle is $U(1)^2$ and we can take $CY_4$ to be a sum of two line bundles over $\mathcal{C}$. That is
\begin{equation}
\begin{array}{ccc}
\mathbb{C}^2 & \hookrightarrow & \mathcal{L}_{+} \oplus \mathcal{L}_{-} \\ & & \downarrow \\ & & \mathcal{C}, \label{UVbundle}
\end{array}
\end{equation}  The degrees of the line bundles are constrained by the vanishing of the first Chern class of the $CY_4$.  The $U(1)^2$ structure group corresponds to the phases of the line bundles and are identified with the $U(1)_+\times U(1)_-$ subgroup of the six-dimensional $SO(5)$ R-symmetry. This is precisely the set-up that captures the abelian twists of the six-dimensional $(2,0)$ SCFT described in section \ref{twist}.  

When the branes are backreacted, the near-horizon region is a warped geometry of the form $AdS_{3}\times_w M_8$, where $M_8$ is some compact eight-manifold.  Now the main challenge boils down to classifying all the possible choices of ${M}_{8}$ which admit the following decomposition:
\begin{equation}
\begin{array}{ccc}
N_4 & \hookrightarrow & M_8 \\ & & \downarrow \\ & & M_{\mathcal{C}}. \label{IRbundle}
\end{array}
\end{equation}
Here the base four-manifold $M_{\mathcal{C}}$ is the IR limit of the complex two-dimensional surface $\mathcal{C}$.  The structure group of the bundles must be $U(1)^2$, and $N_4$ must admit at least a $U(1)^2$ isometry in accordance to the fact that $M_8$ descended from a $CY_4$ which is a sum of two line bundles. 

Now putting everything together, we conclude that in the region near M5-branes, the eleven-dimensional metric must be of the form
\begin{equation}
M^{1,10}\rightarrow AdS_{3} \times_{w} (M_{\mathcal{C}}\times S^{1}_{+}\times S^{1}_{-}\times [t^{+}] \times [t^{-}]) \label{IRspacetime}.
\end{equation}
The  circles dual to $U(1)_{\pm}$ are represented by $S^{1}_{\pm}$; they are generically fibered over the 4d base manifold $M_{\mathcal{C}}$. The last two directions with coordinates $t^{\pm}$ do not, generically, correspond to any symmetries. The $AdS_{3}$ radius and the two interval directions $t^{\pm}$ are combinations of the real line $\mathbb{R}$ in \eqref{UVspacetime}, and the radii of the fibers in the line bundles. In general the metric on ${M}_{8}$ will depend on the coordinates of $M_{\mathcal{C}}$ and on $t^{\pm}$; therefore the restrictions set by this metric ansatz \eqref{IRspacetime} do not seem very constraining.  Nevertheless, we can still gain some control of the supergravity equations and bring them to a manageable form.  At last, in addition to this metric ansatz, we will also need to make an ansatz for the four-form flux.  

In section \ref{02fromM5} we describe the general form of $AdS_3\times_w M_8$ solutions of M-theory from configurations of M5-branes.  In section \ref{ansatz} we impose the $U(1)^2$ isometry for $M_8$ and further reduce the BPS system.  Then, in section \ref{ansatzflux} we fix an ansatz for the four-form flux.  This has the benefit of greatly simplifying the metric and the BPS equations.  Finally in section \ref{02system} we describe the metric for the spacetime in \eqref{IRspacetime} that describes the system of interest.  This last section is a summary of our main results and is written so that it can read separately from the previous ones.  The reader interested merely in the results can directly go to this section.

%%%%%%%%%%%%%%%%%%%%%%%%%%%%%%%%%%%%%%%%&&&&%%%%%%%%
\subsection{$AdS_{3}$ spacetimes from M5-branes}\label{02fromM5}
%%%%%%%%%%%%%%%%%%%%%%%%%%%%%%%%%%%%%%%%%%%%%%%%%%%%%

The general BPS conditions for $AdS_3$ solutions in M-theory are obtained in \cite{Gauntlett:2006qw} by using G-structure analysis.  We describe these results in Appendix \ref{BPSappendix} and reduce them for the case where the solutions arise solely from M5-branes.  In this section we summarize these results.  

The eleven-dimensional metric for gravity duals of $\mathcal{N}=(0,2)$ SCFTs from configurations of M5-branes is given as
\begin{equation}
\begin{split}
ds^{2}_{11}&=L^{4/3} e^{2\lambda}\big[ds^{2}(AdS_{3})+e^{-6\lambda}d\hat s^{2}(M_{8})\big]\\
d\hat s^{2}({M}_{8})&=e^{3\lambda}\hat g^{6d}_{MN}\,dx^{M}dx^{N}+\frac{1}{\text{sin}^{2}2\beta}dy^{2}+\frac{e^{6\lambda}\text{sin}^{2}2\beta}{n^{2}}(d\psi+P)^{2} \label{M8M5system}
\end{split}
\end{equation}
where $\hat g^{6d},\lambda, \beta $ and $P=P_{M}dx^{M}$ are all functions of $x^{M}$ and $y$. The $AdS$ radius is $L$; it can be factored out as an overall parameter.  We also have
\begin{enumerate}
\item $\hat g^{6d}$ is a family of complex metrics on $M_{6}$ parametrized by $y$
\item $\partial_{\psi} $ is a Killing vector and generates a $U(1)_\psi$ that is dual to the $U(1)$ R-symmetry of the $\mathcal{N}=(0,2)$ SCFT$_2$,
\item the corresponding complex structure is independent of $\psi$ and $y$ .
\end{enumerate}

The BPS equations can be written in terms of a compatible pair of $SU(3)$-structure forms on $M_6$; namely an almost symplectic structure, $\hat{J}$, and a holomorphic three-form, $\hat{\Omega}$.  The integrability equations for $\hat{J}$ are given by
\begin{subequations}
\begin{align}\label{Jeq}
i_{\bj}d_{6}\bj&=0\\
i_{\bj}\partial_{y}\bj&=-\frac{1+\text{cos}^{2}2\beta}{\text{sin}^{2}2\beta}\partial_{y} \ln\,e^{3\lambda}\\
\frac{1}{n} i_{\bj}d_{6}P&=2\frac{\cs}{\text{sin}^{2}2\beta}\partial_{y}\ln e^{3\lambda}\\
\partial_{\psi}\bj&=0.  
\end{align}
\end{subequations}
The integrability equations for the holomorphic three-form are given by
\begin{subequations}
\begin{align}
d_{6}\bom&=\Big[\frac{2i}{n}P-\frac{1}{2}d_{6}\ln(\sn\tn)\Big]\wedge\bom\label{d6om}\\
y\partial_{y}\bom&=-\frac{1}{2}\Big[y\partial_{y}\ln(\sn\tn)+\frac{1+\text{cos}^{2}2\beta}{\text{sin}^{2}2\beta}\,\Big]\bom\\
\partial_{\psi}\bom&=i\frac{2}{n}\bom. \label{psicharge}
\end{align}
\end{subequations}  Note that the holomorphic three-form carries a charge under the $U(1)_\psi$ given as $\frac{2}{n}$.  

The four-form flux can be written completely in terms of the warp factor and almost symplectic structure $\hat{J}$; it is given as
\begin{equation}
\begin{split}
B_{4}=&\Big[\partial_{y}\bj+\frac{1}{n}\cs\, d_{6}P\Big]\wedge\bj+\cot 2\beta \star_{6} d_{6} (\sec 2\beta \,\bj)\wedge\hr+\frac{1}{\sn}d_{6}\Big(\cs\bj\Big)\wedge\hp\\
&+\big[2\bj-\star_{6}\frac{1}{n}d_{6}P\wedge\bj-\frac{\cs}{2}\star_{6}\partial_{y}(\bj\wedge\bj)\big]\wedge\hr\wedge\hp, \label{flux}
\end{split}
\end{equation}
with 
\begin{equation}
\cs=2y e^{-3\lambda},\qquad \hr=\frac{e^{-3\lambda}}{\sn}dy,\qquad \hp=\frac{\sn}{n}(d\psi+P) \label{RhoPsiforms}.
\end{equation}
The BPS equations tell us that $B_4$ must satisfy
\begin{align}
(e^{3\lambda}\cos 2\beta)\,  d_8B_4+d_8\left(e^{3\lambda}\star_8B_4\right)=0.
\end{align}  We therefore see that when we impose the Bianchi identity
\begin{equation}
d_{8} B_4 =0,
\end{equation} 
the equation of motion of $B_4$ naturally emerges from the BPS equations.  In other words, solving equations \eqref{Jeq}-\eqref{psicharge} together with the Bianchi identity is equivalent to solving the supergravity equations (see \cite{Gauntlett:2006qw,MacConamhna:2005ip}). Thus, the main strategy we shall follow for finding solutions, is  to first solve the BPS equations for the metric $\hat{g}^{6d}$ and warp factor $e^{3\lambda}$, and further restrict the solutions to those that solve the Einstein equations by imposing the Bianchi identity.  

%%%%%%%%%%%%%%%%%%%%%%%%%%%%%%%%%%%%%%
\subsection{The metric ansatz for $U(1)^2$ systems}\label{ansatz}
%%%%%%%%%%%%%%%%%%%%%%%%%%%%%%%%%%%%%%

In this section, we describe the metric ansatz for the spacetime in \eqref{IRspacetime}.   The geometry of $M_{8}$ in \eqref{M8M5system} has a built-in $U(1)_{\psi}$ isometry and admits the decomposition
\begin{align}
M_8\rightarrow M_{6}\times \tilde S^2_{(y,\psi)},
\end{align} 
where $\tilde S^2_{(y,\psi)}$ is a squashed 2-sphere fibered on some generic complex 6-manifold $M_{6}$. Now in accordance to \eqref{IRspacetime} we need to impose a $U(1)_\phi$ isometry for $\hat g^{6d}$.  This can be done by decomposing $M_6$ as
\begin{align}
M_6\rightarrow M_{\mathcal{C}}\times \tilde S^2_{(z,\phi)},
\end{align} 
where $\tilde S^2_{(z,\phi)}$ is a squashed 2-sphere fibered on the base four-manifold $M_{\mathcal{C}}$. Hence, the most general ansatz for ($\hat g^{6d}, \bj, \bom)$ is given by
\begin{align}
d\hat s^2(M_6) &= e^{2w} ds^2(M_{\mathcal{C}}) + e^{2F-3\lambda} ||\eta_z + i e^C \eta_\phi ||^2 \label{metansb}\\
\hat{J} &= e^{2w} J_{\mathcal{C}} + e^{2F-3\lambda} e^{C} \eta_z \wedge \eta_\phi \label{Jansb} \\ 
\hat{\Omega} &= e^{i\psi + i Q_\phi \phi} e^{2w}e^{F - \frac{3}{2} \lambda} ~\Omega_{\mathcal{C}} \wedge \left(\eta_z + i e^C \eta_\phi \right)  \label{omansb}.
\end{align} This decomposition defines a (local) $SU(2)$ structure on $M_{\mathcal{C}}$; this is defined by a pair of a compatible almost symplectic structure, $J_\mathcal{C}$,  and a holomorphic two-form, $\Omega_\mathcal{C}$, on $M_{\mathcal{C}}$. 

The $\eta$ one-forms are given by
\begin{equation}
\eta_z = dz + V^R, \qquad \eta_\phi = d\phi + V^I,
\end{equation}  where $V^R$ and $V^I$ are some real one-forms on $M_\mathcal{C}$. All forms and functions depend on the coordinates on $M_\mathcal{C}$, $y$ and $z$.  In writing the general ansatz, we also allow for the function $C$.  However, when $C$ is independent of $y$, we can set it to zero by redefining the coordinate $z$ and by shifting the one-form $V^R$ and function $F$.  When we expand the BPS equations we find that $C$ is indeed independent of $y$; we therefore fix  $C$ to zero.  

Without loss of generality, the charge of the holomorphic three-form under $U(1)_\psi$, in equation \eqref{psicharge}, can be fixed as $\frac{2}{n} =1$.  The coordinate $\phi$ parametrizes a circle that is dual to $U(1)_\phi$.  The parameter $Q_\phi$ is the charge of the holomorphic three-form under $U(1)_\phi$; while it is allowed, we can fix it as $Q_\phi =0$ by shifting $\psi$ and the connection one-form $P$.  Since we can make the holomorphic form neutral under $U(1)_\phi$, the killing spinor for the background must be neutral under $U(1)_\phi$; this isometry $\partial_\phi$ is then dual to the flavor symmetry $\mathcal{F}$ in \eqref{U(1)syms}. 

Now notice that so far we have decomposed $M_8$ into a base manifold $M_{\mathcal{C}}$ and two squashed 2-spheres $\tilde S^2_{(z,\phi)}$ and $\tilde S^2_{(y,\psi)}$. In particular, the 2-spheres have some non-trivial composition which defines a four-manifold, parametrized by the coordinates $(z,\phi,y,\psi)$, and this in turn is fibered on the base. This four-manifold is precisely the $N_4$ fiber of the normal bundle 
\begin{equation}
\begin{array}{ccc}
N_4 & \hookrightarrow & M_8 \\ & & \downarrow \\ & & M_{\mathcal{C}}\label {M81}. 
\end{array}
\end{equation}
The metric of the bundle takes the form
\begin{align}
ds^2(M_8)&=e^{2w+3\lambda}ds^2(M_\mathcal{C})+\frac{1}{2}ds^2(N_4),\\
ds^2(N_4)&=2e^{2F}(\eta_z^2+\eta_\phi^2)+\frac{2}{\sin^22\beta}dy^2+\frac{\sin^22\beta}{2}e^{6\lambda}(d\psi+P)^2,\label{M82}
\end{align}
and $ds^2(M_\mathcal{C})$ is some complex metric on $M_\mathcal{C}$ whose choice serves as the "initial data" of the problem. 

At last, recall that earlier we argued that in order to capture the twisted compactifications of the (2,0) theory in supergravity, the structure group of this bundle must be $U(1)^2$. This implies that we must trivialize the connection form $V^R$ appearing in $\eta_z$.  We achieve this in the following way. First, we decompose the exterior derivative on $M_6$ as
\begin{equation}
d_6 = d_{\mathcal{C}} + \eta_z \wedge \partial_z, \qquad d_{\mathcal{C}} \equiv d\hat{x}^{\mu} \partial_\mu  - V^R \wedge \partial_z
\end{equation} where $\hat{x}^\mu$ are coordinates on $M_{\mathcal{C}}$.  The twisted differential $d_{\mathcal{C}}$ is the natural object that appears in the system of equations. Next we assume that $V^R$ is exact and we write
\begin{equation}
V^R = d_{\mathcal{C}} \Gamma,
\end{equation} for some function $\Gamma$ which depends on all the coordinates.  We then consider a coordinate transformation from $(\hat{x}^\mu, y, z)$ to $(x^\mu, y', z')$ defined by
\begin{align}
x^\mu &= \hat{x}^\mu, \qquad y'= y'(y), \qquad z = -\Gamma(x^\mu,y',z'), \\
\text{s.t.}\qquad d_{\mathcal{C}} &= dx^\mu \partial_\mu, \qquad \eta_z = - (\partial_{y'} \Gamma) dy' - (\partial_{z'} \Gamma) dz'.
\end{align}
Hence, after the transformation the connection form $V^R$ gets completely eliminated from $\eta_z$ and in addition, the twisted differential operator $d_\mathcal{C}$ becomes the exterior derivative on $M_\mathcal{C}$.  

Having set up the metric ansatz, we next need to reduce the BPS equations and find the constraints that the metric on $M_\mathcal{C}$, or equivalently, the $SU(2)$ structure $(J_\mathcal{C},\Omega_\mathcal{C})$ must obey.
In appendix B, we discuss this reduction in great detail. In particular, we find that the base four-manifold $M_\mathcal{C}$ is conformally K\"ahler, whose volume depends only on the $x^\mu$ coordinates of $M_\mathcal{C}$. Moreover, we find that the BPS equations reduce significantly the number of unknown metric functions and they partially fix the $U(1)_{\psi/\phi}$ connections $P/V^I$.  However, the resulting set of equations is still highly non-linear and in order to simplify it further we need to impose appropriate constraints on the four-form flux, $B_4$.

%%%%%%%%%%%%%%%%%%%%%%%%%%%%%%%%%%%%%%%%%%%%%%
\subsection{Ansatz for the four-form flux}\label{ansatzflux}
%%%%%%%%%%%%%%%%%%%%%%%%%%%%%%%%%%%%%%%%%%%%%%%%
The system under consideration describes the near-horizon geometry obtained by a stack of $N$ M5-branes wrapped on a complex surface $\mathcal{C}\subset CY_4$. This brane configuration gives rise to a magnetic flux, $B_4$, which lies entirely on $M_8$.  In the region near the branes, the four-form flux encodes the data of the original brane configuration that led to the $AdS_3$ solution.  We must therefore pick an appropriate ansatz for $B_4$ that reflects the M5-brane configurations for the set-up in \eqref{UVbundle}.  
 
The expression of the magnetic flux is given in (\ref{flux}). For the $U(1)^2$ ansatz described above, this expression becomes (schematically) 
\begin{equation}
B_4 = \mathcal{B}^{(0)} \hat{\text{vol}}(N_4) + \frac{1}{3!}\mathcal{B}_{abc}^{(1)} \wedge e^a \wedge e^b \wedge e^c +\frac{1}{2!} \mathcal{B}_{ab}^{(2)} \wedge e^a \wedge e^b + \mathcal{B}_{a}^{(3)} \wedge e^a + \mathcal{B}^{(4)}, \label{B4formal}
\end{equation} where $e^a=\left(dy,\eta_z,\eta_\phi,D\psi\right)$, $\hat{\text{vol}}(N_4)=e^1\wedge e^2\wedge e^3\wedge e^4$ is the unit volume form on $N_4$ and the $\mathcal{B}^{(n)}$ are $n$-forms on the base manifold $M_\mathcal{C}$. Now the number of M5-branes wrapped on the surface $\mathcal{C}$ is measured by the magnetic flux threading a normal four-cycle.  In the near brane region, the flux is expected to lie entirely on the four-cycle defined by $N_4$ and it must therefore asymptote to the $\mathcal{B}^{(0)}$ term in \eqref{B4formal}.  However, $N_4$ is non-trivially fibered over $M_\mathcal{C}$ and the $\mathcal{B}^{(0)}$ term, alone, cannot satisfy the Bianchi identity.  We must thus allow for other terms in \eqref{B4formal} which together with the $\mathcal{B}^{(0)}$ term conspire to the closure of $B_4$. At a minimal level, this is achieved by allowing for $\mathcal{B}^{(1)}$ and $\mathcal{B}^{(2)}$ terms. Having said that, we then need to impose that
 \begin{equation}
 \mathcal{B}^{(4)} =0, \qquad \mathcal{B}^{(3)}_a =0. \label{fluxcons}
 \end{equation} 
 
 Now, these  additional constraints yield useful equations that greatly simplify the metric and the BPS equations.  In appendix \ref{fluxapp}, we thoroughly discuss their consequences. Here it is worth mentioning that upon analyzing these conditions, we find that there exists a natural set of coordinates on the $N_4$ fiber, which makes apparent the $U(1)_+\times U(1)_-$ isometry group coming from the $\mathbb{C}^2$-bundle \eqref{UVbundle}. To understand the physical significance of this result, recall that $N_4$ can be thought of as an $\hat S^2_{(y,\psi)}$-bundle over $\hat S^2_{(z,\phi)}$. Then in this parametrization the 11d metric appears to enjoy a $U(1)_\psi \times U(1)_\phi$ isometry. However, we expect that this isometry group, which we see near the horizon, originated from the $U(1)_+\times U(1)_-$ isometry group appearing in the full M-theory solution. We should therefore consider the coordinate transformation
 \begin{align}
 (y,z,\psi,\phi)&\longrightarrow(t^+,t^-,\phi_+,\phi_-)\\
\text{s.t.} \qquad N_4: \qquad [y]\times [z]\times S^1_\psi\times S^1_\phi& \longrightarrow [t^+]\times [t^-]\times S^1_{+}\times S^1_{-}, \label{IRN4}
\end{align}
where the $S^1_\pm$ circles now give rise to  the $U(1)_\pm$ isometries. Such a transformation can only be defined once we impose the flux constraints \eqref{fluxcons}. As a consequence of this transformation, the one-forms $(\eta_\phi,D\psi)$ transform into a new basis defined as
\begin{align}
\eta_\pm=d\phi_\pm + P_\pm,
\end{align}
where $P_\pm$ are the $U(1)_\pm$ connection forms of the $S^1_\pm$ circles.

%%%%%%%%%%%%%%%%%%%%%%%%%%%%%%%%%%%%%%%%%%%%%%%%%%%%%%%
\subsection{The Metric System}\label{02system}
%%%%%%%%%%%%%%%%%%%%%%%%%%%%%%%%%%%%%%%%%%%%%%%%%%%%%%%%%%%

In sections \ref{ansatz} and \ref{ansatzflux}, we fix a metric ansatz for $AdS_3$ solutions that appear in the near-horizon limit of M5-branes wrapping a four-manifold by considering the symmetries imposed by the bundle in \eqref{UVbundle}.  In this section, we discuss the metric corresponding to these geometries after taking into account the BPS equations described in section \ref{02fromM5}.  

The BPS equations fix the form of the 11d metric to
\begin{align}
ds^2_{11} &= L^{4/3} e^{2\lambda} \left[ ds^2(AdS_3) + e^{2w-3\lambda} ds^2(M_\mathcal{C})+ \frac{1}{2} e^{-6\lambda} ds^2(N_4) \right] \label{metric02}\\
ds^2(N_4) &= g_{ij} dt^i dt^j + 4 h^{ij} \eta_i \eta_j
\end{align} with $i,j\in\{+,-\}$.  The $AdS$ radius is denoted as $L$.  

%%%%%%%%%%%%%%%%%%%%%%%%%%%
\subsubsection*{The $M_{\mathcal{C}}$ base}

The geometry of $M_\mathcal{C}$ is described by an $SU(2)$ structure $(J_\mathcal{C}, \Omega_\mathcal{C})$ which satisfies 
\begin{equation}
d_\mathcal{C} \left(e^{2w -3\lambda} J_\mathcal{C} \right) = 0,\qquad \partial_i \left(\text{vol}_\mathcal{C} \right) =0, \qquad \partial_i \Omega_\mathcal{C} =0, \label{MCcond}
\end{equation}
where $\text{vol}_\mathcal{C}\equiv\frac{1}{2}\jc\wedge\jc$ is the volume form on $M_\mathcal{C}$. The integrability condition of $\Omega_\mathcal{C}$ is given below, \eqref{holom}.  The four-manifold can be chosen so that it is K\"ahler by absorbing the conformal factor $e^{2w-3\lambda}$.  Our choice in \eqref{metric02} and in \eqref{MCcond} is such that the volume of the manifold and the holomorphic two-form are constant on $N_4$.

%%%%%%%%%%%%%%%%%%
\subsubsection*{The $N_4$ fiber}

The $N_4$ fiber is parametrized by the coordinates $\left(t^+,t^-,\phi_+,\phi_-\right)$; these correspond to the $[t^{\pm}]$ intervals and $S^1_\pm$ circles of the spacetime in \eqref{IRspacetime}.  All the metric functions of $ds^2(N_4)$ are determined by a single potential, $D_0$, and are given as
\begin{equation}
g_{ij} = - \partial_i \partial_j D_0, \qquad h_{ij} = - \partial_i \partial_j \left(D_0 + t (\ln t-1) \right), \qquad \text{where}\qquad t= a_+ t^++ a_- t^-.
\end{equation} 
Here, $a_\pm$ are the same parameters that appear in the $\mathcal{N}=(0,2)$ R-symmetry in \eqref{Rsymfield}.  In field theory, the $a_\pm$ parameters are fixed by c-extrimization \cite{Benini:2012cz}; in the gravity duals, they are fixed by the BPS equations.   The matrix $h^{ij}$, appearing in the metric, is the inverse of $h_{ij}$.  

The one-forms $\eta_\pm$, dual to the $S^1_\pm$ circles, can be written as
\begin{equation}
\eta_\pm = d\phi_\pm + \partial_\pm P_0,  
\end{equation}  where $P_0$ is a one-form on $M_\mathcal{C}$ which determines the $U(1)_\pm$ connections of $N_4$, $P_\pm$.  Moreover, the associated Killing vectors  $(\partial_{\phi_+},\partial_{\phi_-})$ are dual to the $(J^+,J^-)$ generators  of  the $U(1)_+ \times U(1)_-$ Cartan group of the 6d R-symmetry.  In the gravity dual picture, the 2d R-symmetry and flavor symmetry, as defined in \eqref{U(1)syms}-\eqref{Rsymfield}, are described by the Killing vectors   
\begin{equation}
R_{\mathcal{N}=(0,2)}: \quad \partial_\psi = a_+ \partial_{\phi_+} + a_- \partial_{\phi_-}, \qquad \mathcal{F}: \quad \partial_\phi = -\frac{1}{2}\left(\partial_{\phi_+} -\partial_{\phi_-}\right).
\end{equation}

%%%%%%%%%%%%%%%%%%
\subsubsection*{The BPS equations}

The equations split into two classes; namely, into a set of holomorphicity conditions and a pair of BPS equations. The natural object appearing in the holomorphicity conditions is the complex one-form
 \begin{equation}
 V_0\equiv P_0 + \frac{i}{2} d_\mathcal{C} D_0.
 \end{equation} This one-form is fixed by the conditions
\begin{equation}\label{holom}
d_\mathcal{C} \Omega_{\mathcal{C}} = \left(\partial_+ + \partial_-\right) \left[ i V_0 \wedge \Omega_\mathcal{C} \right], \qquad \partial_i \left[\Omega_{\mathcal{C}} \wedge d_\mathcal{C} V_0 \right] =0, \qquad \partial_i \partial_j\left[ \Omega_\mathcal{C} \wedge V_0 \right]=0.  
\end{equation}  
The natural objects appearing in the BPS equations are the warp factors $e^{3\lambda}$ and $e^{2w}$. Like the metric on $N_4$, these functions are completely determined by the potential $D_0$; more precisely they are given by
\begin{equation}
 e^{-6\lambda}= \frac{1}{8t}\frac{\det(h)}{\det(g)},\qquad e^{4w} = \frac{t^2}{2}  \det(g) e^{(\partial_+ + \partial_-) D_0}. \label{warpfactors}
\end{equation} 
At last, the BPS equations read as
\begin{align}
\partial_i (e^{4w}) \,\text{vol}_\mathcal{C} &= - e^{2w-3\lambda} J_\mathcal{C} \wedge d_\mathcal{C} \partial_i P_0.  \label{BPSeq}
\end{align}

In addition to the system described here, we also need to impose the Bianchi identity for the four-flux whose reduction is described in appendix \ref{fluxapp} and final form is given in equation \eqref{fluxform}.  It is completely determined by $J_\mathcal{C}$, $P_0$ and $D_0$.  The constraints from the Bianchi identity are significantly more complicated to analyze than the BPS equations.  Therefore it is not very instructive or any useful to write them down in generality.  Nevertheless, if we consider solutions to the BPS equations first, these equations can greatly simplify and this will be our main strategy for constraining the space of solutions of these systems.  

%%%%%%%%%%%%%%%%%%%%%%%%%%%%%%%%%%%%%%%%%%%%%%%%%%%%%%%%%%
\section{Enhanced Supersymmetry}\label{enhancesym}
%%%%%%%%%%%%%%%%%%%%%%%%%%%%%%%%%%%%%%%%%%%%%%%%%%%%%%%

For the class of SCFTs obtained from M5-branes, the supersymmetry can get enhanced for special cases of the normal bundle.  In the field theory section \ref{twist}, we briefly discussed the enhancements to $(0,4)$ and $(2,2)$.  In this section, we discuss these special configurations and reduce the general $(0,2)$ system from section \ref{02system}.  

%%%%%%%%%%%%%%%%%%%%%%%%%%%%%%%%%%%%%%%%%%%%%%%%%%%%%%%%%
\subsection{Systems with $\mathcal{N}=(0,4)$ Supersymmetry}\label{Sys04}
%%%%%%%%%%%%%%%%%%%%%%%%%%%%%%%%%%%%%%%%%%%%%%%%%%%%%%%%%%%

We first consider configurations where the general $\mathcal{N}=(0,2)$ system enhances to $\mathcal{N}=(0,4)$.  In this special case, the surface wrapped by the M5-branes, $\mathcal{C}$, is now embedded in a Calabi-Yau three-fold that is locally $CY_3\cong \mathcal{C} \times \mathbb{C}$.  The eleven-dimensional spacetime of M-theory decomposes as
\begin{equation}
M^{1,10} \rightarrow \mathbb{R}^{1,1} \times CY_3 \times \mathbb{R}^3. \label{04UVbundle}
\end{equation}  The $CY_3$ is the canonical line bundle over $\mathcal{C}$, and the M5-branes are extended along $\mathbb{R}^{1,1} \times \mathcal{C}$.  This configuration is a special case of \eqref{UVbundle} where one of the line bundles $\mathcal{L}_\pm$ is trivial.  It preserves an $SU(2)\times U(1)$ subgroup of the $SO(5)$ R-symmetry of the six-dimensional $(2,0)$ CFT living on the branes.  This system also preserves four supercharges, thus the two-dimensional CFTs in the IR can preserve $\mathcal{N}=(0,4)$ supersymmetry.  In the region near the branes the spacetime decomposes as
\begin{equation}
M^{1,10} \rightarrow AdS_3 \times M_{\mathcal{C}} \times S^1 \times [\tau] \times S^2.   \label{IRSpacetime04}
\end{equation}  The $S^1$ is non-trivially fibered over $M_\mathcal{C}$ and descends from the phase of the line bundle; the $S^2$ corresponds to the isometries of the $\mathbb{R}^3$ factor, in \eqref{04UVbundle}, and is dual to the $SU(2)$ R-symmetry of the $\mathcal{N}=(0,4)$ SCFTs.  The radii of $\mathbb{R}^3$ and the line bundle combine to give the overall $AdS_3$ radius and the interval $\tau$.  

The metric for the spacetime in \eqref{IRSpacetime04} can be obtained by trivializing one of the circle bundles in the general $(0,2)$ systems in section \ref{02system}.  This is consistent with trivializing one of the line bundles in \eqref{UVbundle}.  Without any loss of generality, we shall trivialize $\mathcal{L}_+$; this can be achieved by turning off the connection for the $S_+^1$ circle.  In addition we need to also diagonalize the metric $h_{ij}$ on the $S^1_\pm$ circles since the $U(1)_+$ must enhance to $SU(2)_+$, the R-symmetry of $(0,4)$ SCFTs.  These conditions equivalent to
\begin{equation}
\partial_+ P_0 =0, \qquad \text{and} \qquad h_{+-}=0.  \label{04condition}
\end{equation}  In doing so, we expect the interval $\tau$ and circle $S^1$ in \eqref{IRSpacetime04} to be identified with the interval $t^-$ and circle $S^1_-$ of \eqref{IRspacetime} and \eqref{IRN4}.  Furthermore, the interval $t^+$ and circle $S^1_+$ should combine to make the $S^2$ of \eqref{IRSpacetime04}.  Our goal then is to show that these, indeed, follow from the condition \eqref{04condition}.  

The restriction in \eqref{04condition} and the rightmost holomorphicity condition in \eqref{holom} imply that the potential $D_0$ is separable and can be written as\footnote{To be more precise, these two conditions actually imply that $D_0 = \widehat{D}(\tau, x^\mu) + f(t^+)-t(\ln t-1)+G(x^\mu)t^+$. However the last term can be set to zero by a proper rescaling of $(e^{2w},\jc,\Omega_\mathcal{C})$. }
\begin{equation}
D_0 = \widehat{D}(\tau, x^\mu) + f(t^+)-t(\ln t-1),  
\end{equation}with the identification $\tau=t^-$. Moreover the BPS equation along $\partial_+$ in \eqref{BPSeq} implies that $e^{4w}$ is independent of $t^+$.  This can be used to fix $f(t^+)$; we find 
\begin{align}
e^{\partial_+ f} = c_0 -c_1 t^+, \qquad \mbox{and} \qquad a_+ a_- =0. \label{solf04}
\end{align} There are two solutions to consider, one with $a_-=0$ and the other with $a_+=0$.  Without loss of generality, we can fix the parameters as $c_0=1$ and $c_1 =4$.  

In both cases the four-manifold $M_\mathcal{C}$ is complex and admits an $SU(2)$-structure $(J_\mathcal{C}, \Omega_{\mathcal{C}})$.  The integrability of the complex structure and the holomorphicity conditions for the connection can be expressed as
\begin{equation}
d_\mathcal{C} \Omega_\mathcal{C} = i V_- \wedge \Omega_{\mathcal{C}}, \qquad \Omega_{\mathcal{C}}\wedge d_\mathcal{C} V_- =0, \qquad \Omega_{\mathcal{C}}\wedge \partial_\tau V_- =0 \label{holsol04},
\end{equation} where $V_-=\partial_\tau V_0$. To this end, it is convenient to introduce the function $D\equiv \partial_\tau \widehat{D}$ and connection one-form $P_-\equiv\partial_\tau P_0$. We then have
\begin{equation}
V_- \equiv P_- + \frac{i}{2} d_\mathcal{C} D.
\end{equation}  In the special case where $\Omega_\mathcal{C}$ is closed, the complex one-form $V_-$ is holomorphic.  This implies that we can write
\begin{equation}
\frac{1}{2} \widehat \star_\mathcal{C} d_\mathcal{C} D = e^{2w-3\lambda}\jc \wedge P_- \label{DP04rel}
\end{equation} where $\widehat \star_\mathcal{C}$ is the hodge star on the metric $d\hat s^2(M_\mathcal{C})\equiv e^{2w-3\lambda} ds^2(M_\mathcal{C})$.  This will have interesting consequences for the BPS equations.

We shall now proceed with the reduction of the metric and BPS equations for the two cases in \ref{solf04}.  In the first case $(a_-=0)$, we obtain a class where $N_4$ admits an $S^2$ and we interpret them as gravity duals of $(0,4)$ SCFTs.  In the second case $(a_+=0)$, the $N_4$ admits a shrinking $T^2$ instead and we interpret this class as gravity duals of NS5-branes on a four-manifold.  We end this section with a brief discussion of general $(0,2)$ systems with $SU(2)$ isometry.  

%%%%%%%%%%%%%%%%%%%%%%%%%%%%%%
\subsection*{Gravity duals to $\mathcal{N}=(0,4)$ CFTs}
%%%%%%%%%%%%%%%%%%%%%%%%%%%%%%%%

The gravity duals of two-dimensional CFTs with $\mathcal{N}=(0,4)$ supersymmetry are given by the metric 
\begin{align}
ds^2_{11} =L^{4/3} \left[ ds^2(AdS_3) + \frac{1}{\sqrt{2}}e^{2w}ds^2(M_\mathcal{C}) -\frac{1}{4} \partial_\tau D d\tau^2 -\frac{1}{\partial_\tau D} \left(d\phi_- + P_-\right)^2 + \frac{1}{4} d\Omega_2^2 \right]. \label{sol04}
\end{align} The reduction to this metric corresponds to the solutions in \eqref{solf04} where $a_-=0$.  In this case the warp factor is constant and given as $e^{-6\lambda} = \frac{1}{2}$; we have rescaled the overall $AdS$ radius in writing the metric.  The metric on the two-sphere is 
\begin{equation}
d\Omega_2^2 =d\theta^2 + \sin^2(\theta) d\phi_+^2, \qquad \mbox{with} \qquad 4t^+ = \cos^2(\theta).  
\end{equation} The $SU(2)$ isometry group of this two-sphere is dual to the R-symmetry of the $\mathcal{N}=(0,4)$ CFTs. Notice that $\partial_+$ has been promoted to an isometry. Furthermore, the symplectic structure on $M_\mathcal{C}$ satisfies the integrability condition
 \begin{align}
 \dc\left(e^{2w}\jc\right)=0,\qquad \text{with} \qquad   e^{4w} = -\frac{1}{2} \partial_\tau e^D.
 \end{align}

The leftover BPS equation \eqref{BPSeq} is best discussed once we consider the four-form flux. This reduces to
\begin{align}
B_4  &=- \frac{1}{\sin\theta}\star_6 d_6 \hat{J} \wedge d\theta + d_6 \hat{J} \wedge (\cos \theta d\phi_+) - \frac{1}{\sqrt{2}} \hat{J} \wedge \text{vol}_{S^2},
\end{align}
where $\text{vol}_{S^2}=\sin\theta d\theta \wedge d\phi_+$ is the volume form on the two-sphere. Recall that $\bj$ is the symplectic structure on $M_6$; for the metric \eqref{sol04}, this is given as 
\begin{align}
\bj=&e^{2w}\jc-\frac{1}{\sqrt{2}}d\tau\wedge\eta_- .
\end{align}
Next, to ensure that the BPS system yields solutions to the supergravity equations of motion we also need to impose the Bianchi identity. In doing so we find that for solutions with a non-trivial $U(1)_-$ connection, $\bj$ must be integrable, i.e. $d_6\bj=0$; this then yields the following condition:
\begin{align}
\partial_\tau(e^{2w}\jc)=-\frac{1}{\sqrt{2}}\dc P_- \label{BPSeq05}.
\end{align}
This condition in turn implies that the BPS equation \eqref{BPSeq} is trivially solved and in addition, the flux further reduces to
\begin{align}
B_4  &=  \left(\frac{1}{2}d\tau\wedge\eta_- - \frac{1}{\sqrt{2}}e^{2w}\jc\right) \wedge \text{vol}_{S^2}.
\end{align}
Thus we find that the space of these $\mathcal{N}=(0,4)$ gravity duals is entirely determined by the holomorphicity conditions \eqref{holsol04} and the BPS equation \eqref{BPSeq05}.

It is worth noticing that these solutions come with two distinctive features.  The first is that there is no warping between the $AdS_3$ and the internal space.  The second is that the topology of the $N_4$ is not an $S^4$ since the sphere dual to the R-symmetry is not warped with the rest of the space.  It is not clear why the gravity duals to $(0,4)$ theories should be this way; however it is important to point out that this result is consistent with \cite{Benini:2013cda} and \cite{Gauntlett:2000ng}.  The $AdS_3$ solutions constructed in these papers were obtained by considering $AdS_3$ vacua of seven-dimensional gauged supergravity.  These systems uplift to M-theory as $AdS_3 \times_w M_{\mathcal{C}} \times \widetilde{S}^4$ where $\widetilde{S}^4$ is a squashed four sphere.  Indeed, the authors did not find $AdS_3$ vacua with $\mathcal{N}=(0,4)$ supersymmetry.  
 
In the special case where $\Omega_{\mathcal{C}}$ is closed, we can use the relation in \eqref{DP04rel} to simplify the BPS equation in \eqref{BPSeq} to 
\begin{equation}
\dc\widehat{\star}_\mathcal{C}\dc D=\partial_\tau^2 D\,\, \text{vol}_{\mathcal{C}}
\end{equation} This equation can be seen as a four-dimensional generalization of the $SU(\infty)$ toda equation that governs the gravity dual of $\mathcal{N}=2$ SCFTs in LLM system \cite{Lin:2004nb}.

 %%%%%%%%%%%%%%%%%%%%%%%
\subsection*{$AdS_3\times CY_3\times S^2$} 
%%%%%%%%%%%%%%%%%%%%%%%%%
So far we have been working with the $U(1)^2$ system for the $\mathcal{N}=(0,2)$ theories as derived in section \ref{02system}. We have seen that in the special limit where the $U(1)_-\times U(1)_+$ isometry group enhances to a $U(1)_-\times SU(2)_+$ (see \eqref{04condition}), the BPS system gives rise to $\mathcal{N}=(0,4)$ gravity duals. These solutions take the form $AdS_3\times M_6 \times S^2_{(t^+,\phi_+)}$ with $M_6\cong M_\mathcal{C}\times_w \tilde S^2_{(\tau,\phi_-)}$. After considering the Bianchi identity, we have found that $M_6$ must be K\"ahler, a fact which hints that this particular class of (0,4) solutions may actually fall into a larger and more generic class. 
Thus in this discussion, instead of working with the $U(1)^2$ system,  we directly reduce the more generic (0,2) system, described in section \ref{02fromM5}, for the (0,4) limit. 

In the previous system, we see that the $S^2_{(t^+,\phi_+)}$ is actually spanned by the one-forms $\hr$ and $\hp$ (see \eqref{RhoPsiforms}), with the identification $(y^2=2t^+,\psi=\phi_+)$ and $P=0$. Thus in order to accommodate the $SU(2)$ R-symmetry of the (0,4) theories we set the $U(1)_\psi$ connection to zero and promote $\partial_y$ into a Killing vector field. The $\bj$ equations then imply that $e^{-6\lambda}$ must be constant, which in turn implies that the function $\beta$ depends only on the $y$-coordinate and therefore  defines an angular coordinate for the metric. In order to form a round $S^2_{(y,\psi)}$, the warp factor must be fixed to $e^{-6\lambda}=\frac{1}{2}$. Notice also that $d_6 \bom=0$.

Now, once again we consider the Bianchi identity to find that $d_6\bj=0$. Since both $\bj$ and $\bom$ are closed on $M_6$, we conclude that $M_6$ must be Calabi-Yau.  The magnetic flux, as given in \eqref{flux}, reduces to
\begin{align}
B_4  &=- \frac{1}{\sqrt{2}} \hat{J} \wedge \text{vol}_{S^2},
\end{align}
where $\text{vol}_{S^2}=-d(\cs d\psi)$. Notice that there are no further restrictions from the supersymmetry conditions. 

Putting everything together, we find that the gravity duals of these two-dimensional CFTs with $\mathcal{N}=(0,4)$ supersymmetry are given by the metric 
\begin{align}
ds^2_{11} =&L^{4/3} \left[ ds^2(AdS_3) + \frac{1}{\sqrt{2}}ds^2(CY_3) + \frac{1}{4} d\Omega_2^2 \right]\\
d\Omega_2^2 =&d(2\beta)^2 + \sin^2(2\beta) d\psi^2.
\end{align}
This is the well known $AdS_3\times CY_3\times S^2$ solutions. As discussed in section 6.2 of \cite{Gauntlett:2006fk}, this class of solutions is obtained when, in the near-horizon, the radial coordinate of the $\mathbb{R}^3$ in \eqref{04UVbundle} is identified with the $AdS_3$ one.

 %%%%%%%%%%%%%%%%%%%%%%%
\subsection*{IIB NS5-branes on four-manifold} 
%%%%%%%%%%%%%%%%%%%%%%%%%

The second class of solutions, $a_+=0$ in \eqref{solf04}, leads to metrics where $N_4$ admits a $T^2$.  The metric is given as
\begin{align}
ds^2_{11} &=L^{4/3} e^{2\lambda} \left[ ds^2(AdS_3) + e^{2w - 3\lambda} ds^2(M_\mathcal{C}) + \frac{1}{2} e^{-6\lambda} ds^2(N_4) \right] \label{04NS5}\\
ds^2(N_4) &=  \frac{1-\tau\partial_\tau D}{\tau} d\tau^2 - \frac{4}{\partial_\tau D} \left(d\phi_- + P_-\right)^2 + ds^2(T^2).
\end{align}   The coordinates $(t^+,\phi_+)$ combine to make an $\mathbb{R}^2$ plane which we identify with a shrinking torus.  This implies that the volume of $N_4$ is non-compact.  However, this plane in M-theory can be taken as a $T^2$ with Planck size volume (see section 6.2 of \cite{Gauntlett:2006qw} ).  From this point of view, these are singular solutions of M-theory.  We can then compactify on one circle to IIA supergravity and then T-dualize on the other to IIB supergravity.  

In IIB supergravity, these solutions describe the near-horizon limit of NS5 branes wrapped on a four-manifold, $M_\mathcal{C}$.  This follows from the fact that when we compactify on a circle that is not wrapped by the M5-branes, we obtain NS5 branes in IIA string theory.  The subsequent T-duality maps them to NS5 branes in IIB string theory that are wrapped on $M_\mathcal{C}$.  We explore these IIB systems in future publications.  

Now the warp factors become
\begin{equation}
e^{4w} = 2(1-\tau \partial_\tau D) e^{D}, \qquad e^{-6\lambda} =- \frac{1}{8} \frac{\partial_\tau D}{1-\tau\partial_\tau D}, \qquad D = \partial_\tau \widehat{D}.
\end{equation}
The symplectic structure, $J_\mathcal{C}$, satisfies a conformally K\"ahler condition
\begin{equation}
d_\mathcal{C} \left(e^{2w-3\lambda} J_\mathcal{C} \right) =0.  
\end{equation}
 In addition to the conditions in \eqref{holsol04}, the BPS equation reduces to 
\begin{equation}
\partial_\tau( e^{4w})\,\text{vol}_\mathcal{C} = -e^{2w-3\lambda} J_\mathcal{C} \wedge d_\mathcal{C} P_-.  \label{BPSeqNS5}
\end{equation} 

In the special case where $\Omega_{\mathcal{C}}$ is closed, we can use the relation in \eqref{DP04rel} to simplify the BPS equation in \eqref{BPSeqNS5} to 
\begin{equation}
\frac{1}{4}d_\mathcal{C} \widehat{\star}_\mathcal{C} d_\mathcal{C} D = \partial_\tau^2 (e^{D})\,\text{vol}_\mathcal{C}.
\end{equation} We obtain a four-dimensional generalization of the $SU(\infty)$ Toda equation.    

%%%%%%%%%%%%%%%%%%%%%%%%%%%%%%%%%%%%%%%%%%%%%%%%%%%%%%%%%%%%%%%%%%%
\subsection*{$(0,2)$ theories with $SU(2)$ flavor}\label{flavortwist04}
%%%%%%%%%%%%%%%%%%%%%%%%%%%%%%%%%%%%%%%%%%%%%%%%%%%%%%%%%%%%%%%%%%%%

In general we can obtain systems with $SU(2)$ global symmetry by degenerating the connection along a line on the $(t^+,t^-)$ plane.  We can consider scenarios where
\begin{equation}
\left(n_- \partial_+ - n_+ \partial_-\right) P_0 =0
\end{equation} for some parameters $n_\pm$.  This degeneration implies that the potential $D_0$ can be written as
\begin{equation}
D_0 = \widehat{D}(\tau,x) + T(t^+,t^-), \qquad \tau = n_+ t^+ + n_- t^-.
\end{equation} For generic values where $n_\pm \neq a_\pm$, the solutions with $SU(2)$ isometry map to \eqref{sol04} by a coordinate transformation.  This is due to the fact that there are large diffeomorphisms that rotate the $\pm$ labels in  the general $(0,2)$ system in section \ref{02system}.  The solutions for generic $n_\pm$ can be obtained by making such transformation which redefines $t^\pm$ and then proceed with the $(0,4)$ reduction above.    

When $n_\pm =a_\pm$ the solutions correspond to trivializing the connection, $V^I$, for the $\eta_\phi$ bundle in the ansatz \eqref{metansb}-\eqref{omansb}.  The directions $(\eta_z,\eta_\phi)$ combine to make a round sphere, and the $U(1)$ flavor symmetry in the dual $(0,2)$ theories is enhanced to an $SU(2)$ flavor symmetry.  This reduction can be performed from the general $(0,2)$ system in section \ref{02fromM5}.  

These more general reductions are expected to capture the systems described in sections 6.2 of \cite{Gauntlett:2006fk} and in \cite{Kim:2012ek}.  The solutions in these papers are claimed to be dual to SCFTs with $\mathcal{N}=(0,4)$ supersymmetry.  It is unclear how this enhancement should occur.  A more careful study of these system should resolve this issue.

%%%%%%%%%%%%%%%%%%%%%%%%%%%%%%%%%%%%%%%%%%%%%%%%%%%%%%%%%%%
\subsection{Systems with $\mathcal{N}=(2,2)$ supersymmetry: I}\label{sys22}
%%%%%%%%%%%%%%%%%%%%%%%%%%%%%%%%%%%%%%%%%%%%%%%%%%%%%%%%%%%

In this section, we reduce the general $\mathcal{N}=(0,2)$ system to cases where the supersymmetry enhances to $\mathcal{N}=(2,2)$.  We summarize our results in next section, \ref{system22II}. 

The $(2,2)$ enhancement occurs when the $CY_4$, in \eqref{UVspacetime}, is decomposable as $CY_4 = CY_2^{(1)} \times CY_2^{(2)}$, where each Calabi-Yau two-fold is the cotangent bundle over a two-dimensional Riemann surface $\mathcal{C}_\sigma$ ($\sigma=1,2$).  The M-theory spacetime decomposes as
\begin{equation}
M^{1,10} \rightarrow \mathbb{R}^{1,1} \times \mathbb{R} \times CY_2^{(1)} \times CY_2^{(2)}.
\end{equation} The world volume of the M5-branes is $\mathbb{R}^{1,1} \times \mathcal{C}_1 \times \mathcal{C}_2$.  This configuration is a special case of \eqref{UVbundle} where the complex surface $\mathcal{C}$ admits the product decomposition $\mathcal{C} = \mathcal{C}_1 \times \mathcal{C}_2$.  We identify the two line bundles $\mathcal{L}_{+/-}$ to the canonical bundles over the complex curves $\mathcal{C}_{1/2}$ respectively.
%\begin{equation}
%\begin{array}{ccc}
%\mathbb{C}^{(+)}\times\mathbb{C}^{(-)} & \hookrightarrow & \mathcal{L}_{+} \oplus \mathcal{L}_{-} \\ & &\pi_+\Bigg{\downarrow}\pi_- \\ & & \mathcal{C}_1\times\mathcal{C}_2.\label{(2,2)bundle}
%\end{array}
%\end{equation} 
  This system preserves four supercharges with a $U(1)_+ \times U(1)_-$ R-symmetry and therefore leads to two-dimensional CFTs with $\mathcal{N}=(2,2)$ supersymmetry.  In the region near the branes, the M-theory metric takes the form
\begin{equation}
M^{1,10} \to AdS_3 \times \left(\Sigma_1 \times S^1_+ \right) \times \left(\Sigma_2 \times S^1_-\right) \times [t^+] \times [t^-]. \label{IRspacetime22}
\end{equation} The two dimensional Riemann surface $\Sigma_\sigma$ are the near-horizon limits of $\mathcal{C}_\sigma$.  The four-manifold in the region near the branes is given as $M_\mathcal{C} = \Sigma_1 \times \Sigma_2$.  The circle $S_+^1$ ($S_-^1$) has a non-trivial connection only on the surface $\Sigma_1$ ($\Sigma_2$).  

Now, the product decomposition of $M_\mathcal{C}$ as appeared in \eqref{IRspacetime22} implies that the metric on $M_\mathcal{C}$ can be written as
\begin{align}
\Omega_\mathcal{C} &=  e^{2B} \left(dx_1 + i dy_1 \right) \wedge \left(dx_2 + i dy_2 \right)\\
J_\mathcal{C} &= e^{3\lambda-2w}\left[ e^{2A_1} dx_1 \wedge dy_1  + e^{2A_2} dx_2 \wedge dy_2\right] \\
ds^2(M_\mathcal{C}) &=  e^{3\lambda-2w}\left[e^{2A_1} \left(dx_1^2 + dy_1^2 \right) + e^{2A_2} \left(dx_2^2 + dy_2^2 \right) \right].
\end{align} 
In addition to this decomposition, we must also impose
\begin{equation}
d_1 \partial_- P_0 =0, \qquad d_2 \partial_+ P_0 =0 \label{22condition},
\end{equation} where $d_\sigma$ are the exterior derivatives on the planes of $\Sigma_g$, i.e. $\dc=d_1+d_2$. 
The conditions on $M_\mathcal{C}$ in \eqref{MCcond} and the compatibility condition between $J_\mathcal{C}$ and $\Omega_\mathcal{C}$ imply 
\begin{equation}
d_1 A_2 = d_2 A_1 =0, \qquad \partial_\pm B =0, \qquad e^{2A_1 + 2A_2} = e^{4B} e^{4w -6\lambda}.  \label{MCcond22}
\end{equation}
The constraint in \eqref{22condition} and the $\Omega_\mathcal{C}$ conditions in  \eqref{holom} imply that the one-form, $P_0$, and scalar potential, $D_0$, can be decomposed as\footnote{Our convention for hodge star operator is $\star_a dx_a = -dy_a$ and $\star_a dy_a = dx_a$.}
\begin{align}
P_0 &= \frac{1}{2} \star_1 d_1 D_1(x_1,y_1,t^+)  + \frac{1}{2} \star_2 d_2  D_2(x_2,y_2,t^-) \label{P022decomp}\\
D_0&= D_1(x_1,y_1,t^+) + D_2(x_2,y_2,t^-) + T(t^+,t^-)-2 (t^+ +t^-) B \label{D022decomp}
\end{align} The potential $P_0$ is given up to closed terms.

The functions $D_\sigma$ and $T$ are arbitrary in the decompositions of $P_0$ and $D_0$; they are constrained by the BPS equations in \eqref{BPSeq}.  These equations reduce to
\begin{equation}
\partial_+ e^{4w} e^{4B} = \frac{1}{2} e^{2A_2} \Delta_1 D_+, \qquad \partial_- e^{4w} e^{4B} = \frac{1}{2} e^{2A_1} \Delta_2 D_- \label{BPS22}
\end{equation} where $\Delta_\sigma = \left(\partial^2_{x_\sigma} + \partial^2_{y_\sigma} \right)$, $D_+ = \partial_+ D_1$ and $D_-=\partial_- D_2$.  The warp factors are given as
\begin{align}
e^{4w} e^{4B} &= \frac{t^2}{2} \det(g) e^{(\partial_+ + \partial_-) T} e^{D_+} e^{D_-} \\
e^{2A_1 +2A_2} &= \frac{t}{16}\det(h) e^{(\partial_+ + \partial_-) T} e^{D_+} e^{D_-}. 
\end{align} Notice that the function $B$ does not appear in any of the BPS equations or in the metric; we therefore fix it as $B=0$. 

We have narrowed the system down to three unknown functions; namely, $D_\pm$ and $T$. A carefully analysis of the constraints on $A_\sigma$ (see \eqref{MCcond22}) shows that when $D_\pm$ are generic, we must choose $T=-t(\ln t-1)$ and that $\det(h)$ must be separable between the two Riemann surfaces.  However, these restrictions are found to be incompatible with the BPS equations in \eqref{BPS22}. We therefore need to let $T$ be generic and in addition, we shall let one of the functions $e^{D_\pm}$ be separable in $(x_\sigma, y_\sigma)$ and $t_\pm$. Without loss of generality, we choose to fix $e^{D_+}$ as 
\begin{equation}
e^{D_+} = e^{2A_1^0(x_1,y_1)}, \quad e^{2A_1} = f_1(t^+,t^-) e^{2A_1^0(x_1,y_1)}, \quad \Delta_1 A_1^0 =-\kappa_1 e^{2A_1^0(x_1,y_1)}.
\end{equation} 
The metric along the $(x_1,y_1)$ directions with the conformal factor $e^{2A_1^0}$ is that of a Riemann surface with curvature $\kappa_1$.   The different choices of $\kappa_1 \in \{-1,0,1\}$ correspond to $H_2$, $T^2$ and $S^2$, respectively.  At last, the functions $A_2$ and $w$ are given as
\begin{align}
16 f_1 e^{2A_2} &=  t \det(h) e^{(\partial_+ + \partial_-) T} e^{D_-} \\
2 e^{4w_0} &= t^2 \det(g) e^{(\partial_+ + \partial_-) T} e^{D_-} , \qquad e^{4w} = e^{4w_0} e^{2A_1^0}.
\end{align} 
\begin{comment}
The BPS equations in \eqref{BPS22} reduce to
\begin{equation}
\partial_+ e^{2 w_0} = -\frac{\kappa_1}{2} e^{2A_2}, \qquad \partial_- e^{4w_0} = \frac{1}{4} f_1 \Delta_2 D_-.  
\end{equation}  The constraint on $f_1$ follows from the integrability relations of $e^{4w_0}$.  The only constraint on the function $T$ is $\partial_+ e^{2w_0}$ relation.  Indeed this equation is not enough to fix the function.  This is a case where we expect the Bianchi Identity to provide additional equations to fix $T$.  
\end{comment}

Now, depending on whether $\kappa_1$ vanishes or not, the system of equations reduces differently.  For a vanishing $\kappa_1$ we find that
\begin{equation}
\kappa_1=0: \qquad \Delta_2 D_- = \frac{4}{f_1} \partial_- e^{4w_0}, \qquad \partial_+ e^{4w_0} =0=\partial_+ f_1 =0.
\end{equation} The only constraint on the function $T$ is the $\partial_+ e^{2w_0}$ relation and we expect solutions where $D_-$ can have a non-trivial dependence on $(x_2,y_2,t^-)$.
 If $\kappa_1$ is non-zero and $e^{D_-}$ is not separable, the equations reduce as
\begin{align}
\kappa_1\neq 0: \qquad \Delta_2 D_- &= \frac{1}{g} \partial_- e^{2A_2}, \qquad e^{4w_0} = \frac{c-2\kappa_1 t^+}{4} e^{2A_2} \\
\partial_+ e^{2A_2} &=\partial_+ g = 0, \qquad f_1 = (c-2\kappa_1 t^+) g.  
\end{align} The $e^{4w_0}$ and $\partial_+ e^{2A_2}$ relations yield incompatible constraints for $T$. Therefore solutions where $e^{D_-}$ is not separable exist only when $\kappa_1=0$.  Now we summarize the main results for $(2,2)$ theories.  

%%%%%%%%%%%%%%%%%%%%%%%%%%%%%%%%%%%%%%%%%%%%%%%%%%%%%
\subsection{Systems with $\mathcal{N}=(2,2)$ supersymmetry: II}\label{system22II}
%%%%%%%%%%%%%%%%%%%%%%%%%%%%%%%%%%%%%%%%%%%%%%%%%%%%%%

From the previous analysis in section \ref{sys22}, we observe that the $(2,2)$ theories fit into two classes.  In this section, we present the metrics in each class.  

%%%%%%%%%%%%%%%%%%%%%%%%%%%%%%%%%%%%%%%%%%%%%%
\subsection*{Gravity Duals to $\mathcal{N}=(2,2)$ CFTs}
%%%%%%%%%%%%%%%%%%%%%%%%%%%%%%%%%%%%%%%%%%%%%%

In the first class, the four-manifold $M_\mathcal{C}$ is product of two constant curvature Riemann surfaces.  The eleven-dimensional metric is given as
\begin{align}
ds^2_{11} &= L^{4/3}e^{2\lambda} \left[ ds^2(AdS_3) +\sum_\sigma f_\sigma(t^+,t^-) e^{2A_\sigma^0} \left( dx_\sigma^2 + dy_\sigma^2 \right) + \frac{1}{2} e^{-6\lambda} ds^2(N_4) \right] \\
ds^2(N_4) &= g_{ij} dt^i dt^j + 4h^{ij} \eta_i \eta_j.
\end{align} The conformal factors of the two planes satisfy the Liouville equation:
\begin{equation}
\left(\partial_{x_\sigma}^2 + \partial_{y_\sigma}^2 \right) A_\sigma^0(x_\sigma,y_\sigma)= - \kappa_\sigma e^{2A_\sigma^0(x_\sigma,y_\sigma)}.
\end{equation} These describe Riemann surfaces with curvature $\kappa_\sigma \in \{-1,0,1\}$\footnote{A representative solution to the Liouville equation is
$
e^{A_\sigma^0} = \frac{2}{1+ \kappa_\sigma \left(x_\sigma^2 + y_\sigma^2 \right)}. $} corresponding to $H_2$, $T^2$ and $S^2$.  The $H_2$ can be replaced with $H_2/\Gamma$ to obtain a compact surface with genus $\mathfrak{g}>1$; $\Gamma$ is the Fuschian subgroup of the $PSL(2, \mathbb{R})$ isometry group of the hyperbolic plane.

The metric components along $N_4$, the warp factor $e^{-6\lambda}$ and the radii of the Riemann surfaces, $f_\sigma$, are given in terms of a single function $T(t^+,t^-)$.  They are given by
\begin{equation}
g_{ij} = - \partial_i \partial_j T, \qquad h_{ij} = - \partial_i \partial_j \left(T + t (\ln t -1)\right), \qquad e^{-6\lambda} = \frac{1}{8t} \frac{\det(h)}{\det(g)}
\end{equation} where $t= a_+ t^+ + a_- t^-$ and $a_+ + a_- =1$.  The radii are given as
\begin{equation}
\kappa_1 f_2 = - \partial_+ \left[t^2 \det(g) e^{(\partial_+ + \partial_-) T} \right], \qquad \kappa_2 f_1 = - \partial_- \left[t^2 \det(g) e^{(\partial_+ + \partial_-) T} \right].
\end{equation} There is an additional constraint between the radii which serves as the only constraint on the function $T$, it is given as
\begin{equation}
f_1 f_2 = \frac{t}{16} \det(h) e^{(\partial_+ + \partial_-) T}.  
\end{equation}  This equation is not enough to fix $f_\sigma$ and $a_\pm$.  Additional constraints should be obtained by considering the Bianchi identity of the four-form flux in appendix \ref{fluxapp}.  The $\mathcal{N}=(2,2)$ solutions discussed in \cite{Benini:2013cda} and \cite{Gauntlett:2000ng} correspond to cases when the $f$'s are constant.  In general we expect solutions with non-constant $f_\sigma$.  

The one-forms dual to the circles are given as
\begin{equation}
\eta_+ = d\phi_+ + 2\kappa_1 \left(\mathfrak{g}_1-1\right) V_1, \qquad \eta_- = d\phi_- + 2\kappa_2 \left(\mathfrak{g}_2-1\right) V_2
\end{equation} where  where $\mathfrak{g}_\sigma$ are the genera of the surfaces and connection\footnote{The connection forms can be written in terms of $A_\sigma^0$ as $
2\kappa_\sigma(\mathfrak{g}_\sigma-1) V_\sigma =  \star_\sigma d_\sigma A_\sigma^0. $} forms satisfy $\int_{\Sigma_\sigma} d_\sigma V_\sigma =2\pi$. If we denote the degree of the line bundles $(\mathcal{L}_+,\mathcal{L}_-)$ over $\Sigma_\sigma$ as $(p_\sigma,q_\sigma)$, we then notice that the vanishing of the first Chern class of $CY_4$ implies equation \eqref{twistsigma}.

At last, these solutions fit in a larger class of $(0,2)$ solutions where $M_\mathcal{C}$ is a product of two Riemann surfaces.  We study this class in a separate publication where we also discuss the details the $(2,2)$ solutions including the flux and the constraints from the Bianchi identity.  
%%%%%%%%%%%%%%%%%%%%%%%%%%%%%%%%%%%%%%%%%%%%%%
\subsection*{D3 branes on a Riemann surface}
%%%%%%%%%%%%%%%%%%%%%%%%%%%%%%%%%%%%%%%%%%%%%%

In the second class, the four-manifold is given as $M_\mathcal{C}= \Sigma_g \times T^2$.  We can compactify to IIA supergravity on one circle of the $T^2$ and T-dualize on the other to IIB supergravity.  This family should be considered as gravity duals of D3-branes on a Riemann surface.  These solutions can also be obtained from the $(0,4)$ system in section \ref{Sys04} by taking $M_\mathcal{C} = \Sigma_g \times T^2$.  

We reduce the ansatz in section \ref{sys22} further to
\begin{equation}
e^{D_+} = \mu_1^2 e^{2\nu_+}, \qquad \mu_1^2 = c_1 - 2 t^+, \qquad e^{2A_1} =  e^{2\nu_+}, \qquad e^{2A_2} = -\frac{1}{8} \partial_\tau e^{D}
\end{equation} where $\nu_+$ and $R$ are constant.  In addition, the BPS equations imply
\begin{equation}
a_+ a_- =0, \qquad  \Delta_2 D = 4 a_-^2 \partial_\tau e^{D} - 2\left(c_1 a_+^2 +2a_- \tau\right) \partial_\tau^2 e^{D}.  
\end{equation} The potential $D$ is $\partial_\tau D_2(x_2,y_2,\tau)$ in \eqref{BPS22}, where, in order to make contact with the notation introduced in section \ref{Sys04}, we have identified $t^-$ with $\tau$ .  The solutions further split into two classes, one with $a_+=0$ and the other with $a_-=0$.  We can write the system for each case.

\subsubsection*{Solutions with $a_-=0$}

In the case where $a_-=0$, the metric 
\begin{align}
ds^2 &= L^{4/3}\left[ds^2(AdS_3) +  ds^2(T^2) + \frac{1}{4} d\Omega_2^2 \right. \\
&\left. - \frac{1}{4} \partial_\tau e^{D} \left(dx_2^2 + dy_2^2\right) -\frac{1}{4} \partial_\tau D d\tau^{2} - \frac{1}{\partial_\tau D} \left(d\phi_- + \frac{1}{2} \star_2 d_2 D \right)^2\right].
\end{align} The warp factor, $e^{-6\lambda}$, is constant and the potential $D$ satisfies the $SU(\infty)$ Toda equation:
\begin{equation}
 \Delta_2 D + \partial_\tau^2 e^{D} =0.
\end{equation} In these solutions, the $(\mu_1, \phi_+)$ combine to a form a two-sphere.  The $(x_2,y_2, \tau, \phi_-)$ combine to form a hyper-K\"ahler manifold.  These solutions can be obtained from \eqref{sol04} by imposing $M_\mathcal{C}= \Sigma_g \times T^2$.  

There is a larger class of solutions with $M_\mathcal{C}= \Sigma_g \times T^2$ in the general $(0,2)$ system.  These are governed by Monge-Amp\`ere systems similar to the one obtained in \cite{Bah:2015fwa} for the gravity duals of M5-branes on Riemann surfaces.  We hope to explore these systems in future work.  

\subsubsection*{Solutions with $a_+=0$}

In the case when $a_+=0$, the metric is 
\begin{align}
ds^2_{11} &= L^{4/3} e^{2\lambda} \left[ ds^2(AdS_3) + ds^2(T^2)  - \frac{\partial_\tau e^{D} }{4} \left(dx_2^2 + dy_2^2\right)  + \frac{e^{-6\lambda}}{2}  ds^2(N_4) \right]\\
ds^2{N_4} &= ds^2(T^2) + \frac{\left(1-\tau\partial_\tau D\right)} {\tau}d\tau^{2} -  \frac{1}{\partial_\tau D} \left(d\phi_- + \frac{1}{2} \star_2 d_2 D \right)^2.
\end{align} The $(\mu_1, \phi_+)$ combine to make a torus.  The warp factor is given as
\begin{equation}
e^{-6 \lambda} = -\frac{1}{8} \frac{ \partial_\tau D}{1-\tau\partial_\tau D}
\end{equation} and the potential $D$ satisfies
\begin{equation}
 \Delta_2 D =4\partial_\tau e^{D}  - 4 \tau \partial_\tau^2 e^{D}.
\end{equation} This solution is also a special case of \eqref{04NS5} where we fix $M_\mathcal{C} = \Sigma_g \times T^2$.  This system has a shrinking $T^4$ and therefore have multiple interpretations depending on how we compactify to IIB supergravity.  We explore them also in future publications.

%%%%%%%%%%%%%%%%%%%%%%%%%%%%%%%%%%%%%%%%%%%%%%%%%%%%%%%%%%
\section{Summary and Discussions}\label{conclusion}
%%%%%%%%%%%%%%%%%%%%%%%%%%%%%%%%%%%%%%%%%%%%%%%%%%%%%%%

The main goal of this work has been to obtain the generating system of gravity duals to $\mathcal{N}=(0,2)$ two-dimensional SCFTs, that describe the IR dynamics of M5-branes wrapped on four-manifolds.  We have considered supersymmetric configurations where the four-manifold is a complex surface, $\mathcal{C}$, embedded in a Calabi-Yau four-fold that is a sum of two line-bundles over $\mathcal{C}$.  We have also described the ansatz for the metric and flux for the $AdS_3$ vacua in M-theory that can emerge in the near-horizon limit of the brane configurations, and then reduced the supergravity equation of M-theory by using the general classifications of supersymmetric $AdS_3$ solutions described in \cite{Gauntlett:2006qw}.

After reducing the supergravity equations, we have found that the solutions are generically a warped product of $AdS_3$ with an eight-manifold of the form $M_8 \simeq M_\mathcal{C} \times N_4$.  The four-manifold $M_\mathcal{C}$ is K\"ahler and corresponds to the near-horizon limit of the surface $\mathcal{C}$.  The four dimensional space $N_4$ is non-trivially fibered over $M_\mathcal{C}$ with a $U(1)^2$ structure group and therefore also admits a $U(1)^2$ isometry.  Its metric is given by a single potential, $D_0$, that depends on the coordinates on $M_\mathcal{C}$ and those on $N_4$, while the $U(1)$ connections of the bundle are determined by a single one-form, $P_0$, which lives entirely on $M_\mathcal{C}$.  The BPS equations reduce to holomorphicity constraints on a complex form constructed from $(D_0,P_0)$, and to partial differential equations between $D_0$, $P_0$ and the K\"ahler two-form on $M_\mathcal{C}$.  With appropriate choice of warping between $M_\mathcal{C}$ and $N_4$, we can make the volume form of $M_\mathcal{C}$ and its holomorphic two-form constant on $N_4$.  The generic dependence of the functions between the two four-manifolds will allow us to describe solutions where $\mathcal{C}$ has boundaries.  

With an appropriate choice of the bundle, the two-dimensional SCFTs can preserve $\mathcal{N}= (0,4)$, $\mathcal{N}=(2,2)$ or high supersymmetry.  We have reduced our general result for the gravity duals of $\mathcal{N}=(0,4)$ and $\mathcal{N}=(2,2)$ SCFTs.  In the case of $(0,4)$, the warp factor is constant and we have found that $N_4$ admits an $SU(2)$ isometry, that is dual to the R-symmetry; the two-sphere corresponding to this isometry has constant radius and shrinks nowhere in the geometry.  When the supersymmetry enhances to $\mathcal{N}=(2,2)$, the base manifold, $M_\mathcal{C}$, reduces to a product of two constant curvature Riemann surfaces.  Their radii, in general, depend on the coordinates on $N_4$.  The warp factor and metric functions are given in terms of a single a function which satisfies a Monge-Amp\`ere equation on $M_\mathcal{C}$.  

Now with these results at hand, one can explore the space of two-dimensional SCFTs from M5-branes and classify the allowed boundaries for the M5-branes.  However, before taking such a step, one first needs to classify all the solutions where $M_\mathcal{C}$ is compact, i.e. identify the metrics on $N_4$.  In general, these will be governed by Monge-Amp\`ere equations.  In an upcoming publication, we study these classes and also discuss solutions where $M_\mathcal{C}$ is a product of two Riemann surfaces.  These solutions should include the ones discussed in \cite{Gauntlett:2000ng,Benini:2013cda}. In fact, this will be an important check for our reductions.  

Once these classes of solutions are constructed, one can naturally consider probe branes wrapping supersymmetric cycles in these backgrounds; these will appear as sources in the BPS equations. Backreacting the probes then amounts to considering $AdS_3$ vacua obtained from M5-branes wrapped on four-manifolds with boundaries. From the field theory perspective, these probes correspond to the possible supersymmetric defects of the SCFTs.  We therefore see that by studying the space of solutions of the BPS equations derived in this paper as well as their possible sources one can provide a classification for the $\mathcal{N}=(0,2)$ SCFTs.

From the reduction so far, we observe that the possible sources for the $(2,2)$ case are fairly restricted since $M_\mathcal{C}$ is a product of two constant curvature Riemann surfaces.  Defects can only appear in the supergravity as sources to the Liouville equations that governs the metrics on the Riemann surfaces.  These simply lead to orbifold fixed points on the Riemann surfaces \cite{Gaiotto:2009gz}.  

At last, we have found that for certain classes of solutions, $M_8$ admits torii submanifolds.  In these classes of solutions we expect to compactify to IIA supergravity and then T-dualize to IIB supergravity.  This procedure will yield gravity duals to SCFTs from either D3-branes on Riemann surfaces with punctures or IIB NS5-branes on four-manifolds.  This is an interesting line of research that we hope to explore in the future.

\acknowledgments

We thank Dennis Nemeschansky,  Phil Szepietowski, Nikolay Bobev, Nicholas Warner, Ken Intriligator, Alessandro Tomasiello and Nicholas Halmagyi for useful discussions.  IB is supported in part by the DOE grant DE-SC0011687, ANR grant 08-JCJC-0001-0, and the ERC Starting Grants 240210-String-QCD-BH, and 259133-ObservableString.  VS is supported in part by the DOE grant DE-SC0011687.  IB is grateful for the hospitality and work space provided by the UCSD Physics Department.  IB gratefully acknowledges support from the Simons Center for Geometry and Physics, Stony Brook University for the Simons Center Summer Workshop 2015.

%%%%%%%%%%%%%%%%%%%%%%%%%%%%%%%%%
\appendix
%%%%%%%%%%%%%%%%%%%%%%%%%%%%%%%%%%

%%%%%%%%%%%%%%%%%%%%%%%%%%%%%%%%%%%
\section{The BPS equations}\label{BPSappendix}
%%%%%%%%%%%%%%%%%%%%%%%%%%%%%%%%%%%%%

In this appendix we discuss the reduction of the BPS equations for $AdS_3$ spacetimes in M-theory. Since we are interested in geometries which are sourced solely by M5-branes, we set the electric flux to zero. Now
the metric for the most general supersymmetric $AdS_{3}$ solution in M-theory has the form
\begin{align}
ds^{2}_{11}=&e^{2\lambda}[ds^{2}(AdS_{3})+ds^2(M_8)]\\
ds^2(M_8)=&ds^{2}(M_{6})+\hat \rho^{2}+\hat \psi^{2}
\end{align} 
where the 6d metric, $ds^{2}(M_{6})$, admits a local $SU(3)$-structure and $\hat
\rho$ and $\hat \psi$ are two unit one-forms which, generically, have some non-trivial dependence on ${M}_{6}$. The SUSY conditions
tell us that the geometry of the overall 8d internal manifold ${M}_{8}$
must satisfy the following differential conditions:
\begin{subequations}
\begin{align}
d\ln e^{6\lambda}&=e^{-3\lambda}\sn \, i_{J\wedge\hat \psi}\, \star_{8}B_{4} \label{J1}\\
d_{8}(e^{3\lambda}\text{cos}\,2\beta)&=2e^{3\lambda}\sn\,\hat \rho\label{rho}\\
d_{8}(e^{3\lambda}\,\sn\,\hr)&=0\\
d_{8}(e^{6\lambda}\,\text{sin}\,2\beta\,\Omega_{I})&=-2e^{6\lambda}(\Omega_{R}\wedge
\hat \psi-\text{cos}\,2\beta\,\Omega_{I}\wedge \hat \rho)\label{omega} \\
d_{8}[e^{6\lambda}(\frac{1}{2}J\wedge J+\cs& \,J\wedge\hat \psi\wedge \hat\rho)]=e^{3\lambda}\sn
\hr\wedge B_{4} \label{Jeq1}\\
d_{8}(e^{6\lambda}\sn\, J\wedge \hat\psi)=2e^{6\lambda}&(\frac{1}{2}J\wedge J+\cs
\,J\wedge\hat\psi\wedge \hat\rho)-e^{3\lambda}(\star_{8}B_{4}+\cs\,B_{4}) \label{Jeq2}
\end{align}
\end{subequations}
as well as the algebraic constraints:
\begin{subequations}
\begin{align}
\Omega_{I}\wedge B_{4}&=0\\
\hp\wedge \Omega_{R}\wedge B_{4}&=0\\
(i_{J\wedge J}+2 \cos 2\beta \,i_{J\wedge \psi \wedge \hr})\star_{8}B_{4}&=12e^{3\lambda} \label{J2}.
\end{align}
\end{subequations}
Here, $B_{4}$ is the magnetic flux on ${M}_{8}$, $\beta$ is some unknown function and $J$ and $\Omega=\Omega_{R}+i\Omega_{I}$ are the (1,1) symplectic and (3,0) holomorphic forms on ${M}_{6}$, respectively. Given a frame $\{e^{i};i=1,\dots ,6\}$ on ${M}_{6}$, we define the $SU(3)$-structure forms by
\begin{equation}
\begin{split}
J&=e^{1}\wedge e^{2}+e^{3}\wedge e^{4}+e^{5}\wedge e^{6}\\
\Omega&=(e^{1}+ie^{2})\wedge(e^{3}+ie^{4})\wedge(e^{5}+ie^{6}).
\end{split}
\end{equation}
This $SU(3)$-structure has originated by reducing an $SU(4)$-structure on ${M}_{8}$. Upon such a reduction we find that $J$ and $\Omega$ satisfy
\begin{equation}
i_{\hr}J=i_{\hp}J=i_{\hr}\Omega=i_{\hp}\Omega=0.
\end{equation}

In addition to the BPS equations, we also need to impose the Bianchi identity
\begin{align}
d_8B_4=0.
\end{align}
Then solving the BPS equations together with the Bianchi identity is equivalent to solving the supergravity equations. For example notice that when we hit (\ref{Jeq2}) with $d_8$ and substitute in \eqref{rho} and (\ref{Jeq1}) we find that\begin{align}
(e^{3\lambda}\cos 2\beta)\,  d_8B_4+d_8\left(e^{3\lambda}\star_8B_4\right)=0.
\end{align}  We therefore see that when we impose the Bianchi identity the e.o.m. of $B_4$ naturally emerges from the BPS equations.

Our goal here is to bring this highly non-trivial set of equations down to a clean form where it is transparent how the various metric functions are governed by this system. We begin our analysis by fixing the conventions which we will be using throughout this reduction. We then consider some (minor) assumptions which are necessary, not only for simplifying the system, but also for accommodating in our $AdS_{3}$ solutions the various ingredients required for studying the dual field theories. Next, we decouple the magnetic flux from the SUSY equations and thereafter we reduce the system to a set of differential equations for the $SU(3)$-structure $(J,\Omega)$.

%%%%%%%%%%%%%%%%%%%%%%%%%%%%%
\subsection*{Conventions}
%%%%%%%%%%%%%%%%%%%%%%%%%%%%%

The orientation on ${M}_{8}$ is given by $\text{vol}_{8}=\frac{1}{6}J\wedge J\wedge J\wedge\hr\wedge\hp$.  We thus split the
star operator as $\star_{8}=\star_{6}\star_{2}$ where $\star_6$ acts on differential $n$-forms $A_{n}$ lying entirely on ${M}_{6}$
and $\star_2$ acts on the 2d space spanned by $\hr$ and $\hp$. It then follows that
\begin{equation}
\begin{split}
\star_{8}A_{4}&=\star_{6}A_{4}\wedge\hr\wedge\hp\\
\star_{8}A_{3}\wedge\hr&=\star_{6}A_{3}\wedge\hp\\
\star_{8}A_{3}\wedge\hp&=-\star_{6}A_{3}\wedge\hr\\
\star_{8}A_{2}\wedge\hr\wedge\hp&=\star_{6}A_{2}.
\end{split}
\end{equation} 
For example
\begin{equation}
\star_{6}J=\frac{1}{2}J\wedge J.
\end{equation}
Given an $n$-form $A_{n}$ and an $(n+m)$-form $B_{{n+m}}$ we define their interior product by 
\begin{equation}
(i_{A}B)_{m}=\frac{1}{m!}(i_{A}B)_{\mu_{1}\dots\mu_{m}}dx^{\mu_{1}}\wedge\dots\wedge dx^{\mu_{m}}
\end{equation}
where
\begin{equation}
(i_{A}B)_{\mu_{1}\dots\mu_{m}}=\frac{1}{n!}A^{\rho_{1}\dots\rho_{n}}B_{\rho_{1}\dots\rho_{n}\mu_{1}\dots\mu_{m}}.
\end{equation}
It then follows that
\begin{subequations}
\begin{align}
i_{fA_{n}+gB_{m}}=&f\times i_{A_{n}}+g\times i_{B_{m}}\\
i_{A_{n}\wedge B_{m}}C_{p}=&i_{B_{m}}(i_{A_{n}}C_{p})\\
i_{X_{1}\wedge Y_{1}}(A_{n}\wedge B_{m})=&i_{X_{1}\wedge Y_{1}}A_{n}\wedge B_{m}+A_{n}\wedge i_{X_{1}\wedge Y_{1}}B_{m}\nonumber\\
&+(-1)^{n-1}\Big(i_{X_{1}}A_{n}\wedge i_{Y_{1}}B_{m}-i_{Y_{1}}A_{n}\wedge i_{X_{1}}B_{m}\Big)
\end{align}
\end{subequations} 
for some functions $f$ and $g$, and some one-forms $X_{1}$ and $Y_{1}$.
In our case, it is useful to consider the following contractions:
\begin{equation}
i_{J}J=3;\quad i_{J\wedge J}J\wedge J=12;\quad i_{J}X_{1}\wedge J=2X_{1}\label{iJX}.
\end{equation}

At last, notice that if we rescale the frame on ${M}_{6}$ by a factor of $e^{3\lambda/2}$ such that $J=e^{-3\lambda}\bj$, we have
\begin{equation}
\star_{6}\bj=\frac{1}{2}e^{-3\lambda} \bj\wedge \bj;\quad      i_{J}=e^{3\lambda}\times i_{\bj}.
\end{equation}

%%%%%%%%%%%%%%%%%%%%%%%%%%%%%%%%
\subsection*{The simplifying ansatz}
%%%%%%%%%%%%%%%%%%%%%%%%%%%%%%%%%%

We begin by specializing the structure of the eight-manifold ${M}_{8}$.  In particular, we expect that $M_8$ naturally admits a $U(1)_\psi$ isometry associated to the R-symmetry of the dual field theory. To accommodate this feature of $M_8$ we shall let $\hp$ be the one-form dual to the $U(1)_\psi$ isometry. Futhermore, this one-form combines with $\hr$ to form a squashed 2-sphere $\tilde S^2_{(y,\psi)}$ which is fibered on $M_6$. 

To be more concrete, let $x^{M}=(x^{\mu},z,\phi)$ be the coordinates on ${M}_{6}$ and $(y,\psi)$ be the coordinates on $\tilde S^{2}_{(y,\psi)}$. 
In these coordinates, we decompose the 8d derivative as
\begin{equation}
d_{8}=d_{6}+dy\wedge \partial_{y}+d\psi\wedge \partial_{\psi}.
\end{equation}
Next we integrate equations (\ref{rho}-c) to find that
\begin{equation}
\hr=\frac{1}{e^{3\lambda}\sn}dy\quad \text{and} \quad e^{3\lambda}\cs=2y.
\end{equation}
Now, we write $\hp$ as
\begin{equation}
\hp=\frac{\sn}{n}(d\psi+P),
\end{equation}
where $P=P_M(x^M,y) dx^{M}$ is the $U(1)_{\psi}$ connection and $n$ is an integer to be fixed. At last, it is convenient to rescale the metric on ${M}_{6}$ such that
\begin{equation}
g^{6d}=e^{-3\lambda}\hat g^{6d};\quad J=e^{-3\lambda}\bj ;\quad \Omega=e^{-9\lambda/2}\bom.
\end{equation}
Putting everything together, we find that the 11d metric reads as
\begin{equation}
\begin{split}
ds^{2}_{11}&=e^{2\lambda}\big[ds^{2}(AdS_{3})+e^{-6\lambda}d\hat s^{2}(M_{8})\big]\\
d\hat s^{2}({M}_{8})&=e^{3\lambda}\hat g^{6d}_{MN}\,dx^{M}dx^{N}+\frac{1}{\text{sin}^{2}2\beta}dy^{2}+\frac{e^{6\lambda}\text{sin}^{2}2\beta}{n^{2}}(d\psi+P)^{2}. 
\end{split}
\end{equation}

%%%%%%%%%%%%%%%%%%%%%%%%%%%%%%%%%%%
\subsection*{The $\bom$ equations}
%%%%%%%%%%%%%%%%%%%%%%%%%%%%%%%%%%%%%%

Recall that by construction we have $i_{\hp}\bj=i_{\hp}\bom=0$. Let $\partial_{\psi}$ be the vector field dual to $\hp$. Then to promote $\partial_{\psi}$ into a Killing vector field we need the $SU(3)$-structures to satisfy
\begin{equation}
\partial_{\psi}\hat J=0; \qquad \partial_{\psi}\bom=iQ_{\psi}\bom,
\end{equation}
where $Q_{\psi}$ is the $U(1)_{\psi}$ charge of $\bom$. We therefore have the decomposition
\begin{equation}
\bom=e^{iQ_{\psi}\psi} \bom_{o}
\end{equation}
for some $\psi$-independent complex (3,0)-form $\bom_{o}$.
Now we expand (\ref{omega}) and collect the $\text{cos}(Q_{\psi}\psi)$ and $\text{sin}(Q_{\psi}\psi)$ terms - these must vanish independently. This will then yield the derivatives of the $\bom_{o,R/I}$ which we reassemble to obtain
\begin{subequations}
\begin{align}
d_{6}\bom&=\Big[\frac{2i}{n}P-\frac{1}{2}d_{6}\ln(\sn\tn)\Big]\wedge\bom\label{d6om}\\
y\partial_{y}\bom&=-\frac{1}{2}\Big[y\partial_{y}\ln(\sn\tn)+\frac{1+\text{cos}^{2}2\beta}{\text{sin}^{2}2\beta}\,\Big]\bom\\
\partial_{\psi}\bom&=i\frac{2}{n}\bom.
\end{align}
\end{subequations}
Note that supersymmetry fixes $Q_{\psi}$  to $2/n$.

%%%%%%%%%%%%%%%%%%%%%%%%%%%%%%%%%
\subsection*{Fixing $B_{4}$}
%%%%%%%%%%%%%%%%%%%%%%%%%%%%%%%%%%%%

The most general form which $B_{4}$ can take is
\begin{equation}
B_{4}=A_{4}+A_{3}\wedge \hr +A'_{3}\wedge \hp + A_{2}\wedge\hr\wedge \hp, 
\end{equation}
where $A_{n}$ is an $n$-form with legs on ${M}_{6}$ only. Substituting into equation (\ref{Jeq1}) and expanding we deduce that
\begin{equation} \label{dJ}
d_{6}(\bj\wedge\bj)=0
\end{equation}
and that
\begin{equation} \label{B4}
B_{4}=\Big[\partial_{y}\bj+\frac{1}{n}\cs\, d_{6}P\Big]\wedge\bj+\frac{1}{\sn}d_{6}\Big(\cs\bj\Big)\wedge\hp+A_{3}\wedge\hr+A_{2}\wedge
\hr\wedge\hp.
\end{equation}
Here we have used the fact that $\partial_{\psi}\bj=0$. Now substituting the above expression for $B_{4}$ into (\ref{Jeq2}) and solving for $\star_{8}B_{4}$
we find that
\begin{equation}
\begin{split}
\star_{8}B_{4}=-\Big[2\cs\,\bj+\partial_{y}\Big(e^{3\lambda}(\sn)^{2}\bj\Big)+\cs\,A_{2}\Big]\wedge\hr\wedge\hp\,\\
+\Big(\frac{1}{n}e^{3\lambda}(\sn)^{3}\bj\wedge\partial_{y}P-\cs\,A_{3}\Big)\wedge\hr\,\,\\
+\Big(e^{-3\lambda}\bj-\frac{1}{n}\,d_{6}P-\cs\,\partial_{y}\bj\Big)\wedge\bj\,\,\\
-\ct\,d_{6}\Big(\scc\,\bj\Big)\wedge\hp.
\end{split}
\end{equation}

Next, we \textit{directly} compute $\star_{8}B_{4}$ by considering the Hodge-dual of (\ref{B4}), and
thereafter, we compare it to the above expression. In doing so, we find that the unknown forms $A_{2}
$ and $A_{3}$ are given by
\begin{align}
A_2=&2\bj- \star_6 \left[\frac{1}{n}d_6P\wedge\bj+\cs\,\partial_y\left(\frac{\bj\wedge\bj}{2}\right)\right]\\
A_3=&\ct\star_6d_6\left(\sec 2\beta\bj\right)
\end{align}
This completely fixes $B_4$. We also find two consistency conditions which read as
\begin{align}
\partial_{y}\left(\frac{\bj\wedge\bj}{2}\right)+e^{3\lambda}\star_{6}\partial_{y}\bj=&-\frac{1}{y}\left[y\partial_{y}\ln\left(\sin2\beta\tan2\beta\right)+\frac{1+\cos^22\beta}{\sin^22\beta}\right]\frac{\bj\wedge\bj}{2}\label{cc1}\\
e^{3\lambda}\sin^42\beta~\bj\wedge\frac{1}{n}\partial_{y}P=&-2\star_{6}d_{6}\left(\cs\right)\wedge\bj\label{cc2}.
\end{align}
 It is also useful to write down $\star_{8}B_{4}$; this turns out to be
\begin{equation}
\begin{split} \label{starB}
\star_{8}B_{4}=\star_{6}\Bigg[\Big(\partial_{y}\bj+\frac{1}{n}\cs\, d_{6}P\Big)\wedge\bj\Bigg]\wedge\hr&\wedge\hp\\
-\frac{1}{\sn}\star_{6}d_{6}\Big(\cs\bj\Big)&\wedge\hr\\
+\Big(\omega^{-3}\bj-\frac{1}{n}\,d_{6}P-\cs\,\partial_{y}\bj\Big)&\wedge\bj\\
-\ct\,d_{6}\Big(\scc\,\bj\Big)&\wedge\hp.
\end{split}
\end{equation}

%%%%%%%%%%%%%%%%%%%%%%%%%%%%%%%%%%%%%%
\subsection*{The $\bj$ equations}
%%%%%%%%%%%%%%%%%%%%%%%%%%%%%%%%%%%%%%%

First of all we observe that $\bj$ and $B_{4}$ appear together in four different SUSY equations. We have already used two of those to fix the magnetic flux. It is then a straightforward exercise to decouple $B_{4}$ from the remaining two and obtain the differential constraints on $\bj$. Using (\ref{starB}), we project  $\star_{8}B_{4}$ onto the $\bj\wedge\hp$ and $\bj\wedge\bj$ directions and then plug into equations $(\ref{J1})$ and $(\ref{J2})$ to obtain
\begin{subequations}
\begin{align}
i_{\bj}\Big(d_{6}\ln\omega^{3}\wedge\bj+d_{6}\bj\Big)&=2d_{6}\ln e^{3\lambda}\\
i_{\bj}\star_{6}\left[\partial_{y}\left(\frac{\bj\wedge\bj}{2}\right)+\frac{1}{n}\cs\, d_{6}P\wedge\bj\right]&=-2\omega^{3}\partial_{y}\ln e^{3\lambda}\\
i_{\bj\wedge\bj}\left[\frac{1}{n}\,d_{6}P\wedge\bj+\cs\,\partial_{y}\left(\frac{\bj\wedge\bj}{2}\right)\right]&=4\cs\,\partial_{y}\ln e^{3\lambda}.
\end{align}
\end{subequations}
Observe that after applying $(\ref{iJX})$, the first equation simplifies to $i_{\bj}d_{6}\bj=0$ which is equivalent to $d_{6}(\bj\wedge\bj)=0$, equation $(\ref{dJ})$. The last two equations involve only the $\partial_{y}$ derivative of $\bj$ and we shall therefore refer to them as the "$\partial_{y}\bj$" equations.

Note that even in the absence of $B_{4}$ these equations still look messy. We can clean them up a bit by the means of the following trick. 
First we observe that if we write $\bj\equiv \sum_{i=1}^{3}\bj_{i}$, with $\bj_1=e^1\wedge e^2$ and etc, then the most general form that $\partial_{y}\bj$  and $d_{6}P$ admit is
\begin{equation}
\partial_{y}\bj=\sum_{i=1}^{3}h_{i}\bj_{i}+U_2^j;\quad \frac{1}{n}d_{6}P=\sum_{i=1}^{3}f_{i}\bj_{i}+U_2^p.
\end{equation} 
Here the $U_2$'s are some two-forms which are not proportional to any of the $\bj_{i}$, i.e. $i_{\bj_i}U_2=0$. Notice that this decomposition makes the contraction with $i_{\bj}$ more transparent; for example
\begin{align}
i_{\bj}\partial_y\bj=\sum_i h_i.
\end{align}
Now, we compute the following quantities
\begin{equation}
\begin{split}
i_{\bj}\star_{6}(\partial_{y}\bj\wedge\bj)=2e^{3\lambda}\,i_{\bj}\partial_{y}\bj;&\quad i_{\bj\wedge\bj}(\partial_{y}\bj\wedge\bj)=4\,i_{\bj}\partial_{y}\bj\\
i_{\bj}\star_{6}(\frac{1}{n}d_{6}P\wedge\bj)=2e^{3\lambda}\,\frac{1}{n} i_{\bj}d_{6}P;&\quad i_{\bj\wedge\bj}(\frac{1}{n}d_{6}P\wedge\bj)=4\,\frac{1}{n} i_{\bj}d_{6}P.
\end{split}
\end{equation}
Plugging into the $\partial_{y}\bj$ equations we find the following set of equations
\begin{equation}
\begin{split}
i_{\bj}\partial_{y}\bj+\cs \,\frac{1}{n} i_{\bj}d_{6}P&=-\partial_{y} \ln e^{3\lambda}\label{dyJ1}\\
\cs \,i_{\bj}\partial_{y}\bj+ \frac{1}{n} i_{\bj}d_{6}P&=\cs\partial_{y} \ln e^{3\lambda}
\end{split}
\end{equation}
which can be solved simultaneously to give
\begin{equation}
i_{\bj}\partial_{y}\bj=-\frac{1+\text{cos}^{2}2\beta}{\text{sin}^{2}2\beta}\partial_{y} \ln e^{3\lambda};\quad \frac{1}{n} i_{\bj}d_{6}P=2\frac{\cs}{\text{sin}^{2}2\beta}\partial_{y}\ln e^{3\lambda}\label{d6P}.
\end{equation}

%%%%%%%%%%%%%%%%%%%%%%%%%%%%%%%%%%%%%%%%%%%%%%
\subsection*{Torsion classes and structure equations}
%%%%%%%%%%%%%%%%%%%%%%%%%%%%%%%%%%%%%%%%%%%%%%

Given an $SU(3)$-structure we can decompose $(d_{6}\bj,d_{6}\bom)$ in terms of $SU(3)$-representations as follows: 
\begin{equation}
\begin{split}
d_{6}\bj &=-\frac{3}{2}Im(\bar W_{1} \bom)+W_{4}\wedge \bj + W_{3}\\
d_{6}\bom&=W_{1}\bj\wedge \bj+W_{2}\wedge \bj+\bar W_{5} \wedge \bom,
\end{split}
\end{equation}
where the $W_{i}$ are the \textit{torsion classes} of the $SU(3)$-structure. For the BPS system under consideration, we see, by inspection, that $W_{1}$ and $W_{2}$ vanish. As a consequence, the almost complex structure defined by $\bom$ is integrable on ${M}_{6}$, i.e. $M_{6}$ is a complex manifold. Moreover, $W_{4}$ is defined as $W_{4}=\frac{1}{2}i_{\bj}d_{6}\bj$ and it vanishes in accordance to the $d_{6}\bj$ equation. Notice that supersymmetry says nothing about $W_{3}$.

%%%%%%%%%%%%%%%%%%%%%%%%%%%%%%%%%%%%
\section{The $U(1)_\phi$ isometry}
%%%%%%%%%%%%%%%%%%%%%%%%%%%%%%%%%%%%

In this appendix we work out the details for the construction of the $U(1)_\phi$ isometry. In particular, we let the complex manifold $M_6$ be an $\tilde S^2_{(z,\phi)}$-bundle over a four-manifold $M_\mathcal{C}$ and we write the ansatz
\begin{align}
d\hat s^2(M_6) &= e^{2w} ds^2(M_{\mathcal{C}}) + e^{2F-3\lambda} ||\eta_z + i e^C \eta_\phi ||^2 \label{metans}\\
\hat{J} &= e^{2w} J_{\mathcal{C}} + e^{2F-3\lambda} e^{C} \eta_z \wedge \eta_\phi \label{Jans} \\ 
\hat{\Omega} &= e^{i\psi + i Q_\phi \phi} e^{2w} e^{F - \frac{3}{2} \lambda}~\Omega_{\mathcal{C}} \wedge \left(\eta_z + i e^C \eta_\phi \right) \label{omans},
\end{align} 
Notice that we have introduced a rescaled metric on $M_\mathcal{C}$. As we will see, the BPS equations allow us to choose the rescaling factor $e^{2w}$ to be such that the volume of $M_\mathcal{C}$ is $y$-independent.
The $\eta$ one-forms are defined as
\begin{align}
\eta_z=dz+V^R,\qquad \eta_\phi=d\phi+V^I.
\end{align}
Here $V^R$ and $V^I$ are real one-forms on the base which depend on all the coordinates. Together they define the complex one-form
\begin{align}
V\equiv V^R+iV^I.
\end{align}
At last, we decompose the exterior derivative on $M_6$ as
\begin{equation}
d_6 = d_{\mathcal{C}} + \eta_z \wedge \partial_z, \qquad d_{\mathcal{C}} \equiv d\hat{x}^{\mu} \partial_\mu  - V^R \wedge \partial_z
\end{equation} where $\hat{x}^\mu$ are coordinates on $M_{\mathcal{C}}$. 

To this end, it is convenient to introduce the function $\Lambda$ defined through
\begin{equation}
\frac{1+\cos^2(2\beta)}{\sin^2(2\beta)}\equiv -(1+y\partial_y \Lambda).
\end{equation} 
Notice that $\Lambda$ is actually defined up to an arbitrary function which is independent of $y$. Now using this definition, we can rewrite the $\partial_y$ equations as
\begin{align}
i_{\hat{J}} y\partial_y \hat{J} = &y \partial_y \ln \left(y\frac{\cos(2\beta)}{\sin^2(2\beta)} e^{\Lambda}\right)\\
y\partial_y \hat{\Omega} = &- \frac{1}{2} y\partial_y \ln \left(\frac{1}{y}\frac{\sin^2(2\beta)}{\cos(2\beta)} e^{-\Lambda} \right) \hat{\Omega}.
\end{align}
When we substitute in the ansatz we find that if we choose the rescaling factor to be
\begin{align}
e^{4w}=\frac{y^2}{8\sin^2 2\beta}e^{\Lambda-2F},
\end{align}
the $\partial_y \bj$ equation yields
\begin{align}
\partial_y \left(J_\mathcal{C} \wedge J_\mathcal{C} \right)=0.
\end{align}
When we consider the $\partial_y \bom$ we obtain
\begin{align}
\partial_y e^C =0,\qquad  \partial_y \Omega_\mathcal{C} =0, \qquad \Omega_\mathcal{C} \wedge \partial_y \left(V\right ) =0.
\end{align}
Now since $C$ is independent of $y$, we can set it to zero by redefining the coordinate $z$ and by shifting $F$.  From now on, we assume that $C$ is zero.

Next we want to reduce the $d_6\bom$ equation. For, we decompose the $U(1)_\psi$ connection as 
\begin{equation}
\frac{2}{n} P = \frac{1}{2}P_\mathcal{C} +p_z \eta_z+ p_\phi \eta_\phi
\end{equation} where $P_\mathcal{C}$ is some real one-form on $M_\mathcal{C}$ and $p_z$ and $p_\phi$ are some real functions. Equation \eqref{d6om} then reduces to  
\begin{equation}
d_\mathcal{C} \Omega_\mathcal{C} =  \left(i P_\mathcal{C} - \frac{1}{2} d_\mathcal{C} \Lambda \right) \wedge \Omega_\mathcal{C}, \qquad \partial_z \Omega_\mathcal{C}=(ip_z-p_\phi-\frac{1}{2}\partial_z\Lambda)\Omega_\mathcal{C}, \qquad \Omega_\mathcal{C} \wedge d_\mathcal{C} V = \Omega_\mathcal{C} \wedge \partial_z V =0.  
\end{equation}
The integrability condition $[\partial_y,\partial_z]\Omega_\mathcal{C}=0$ fixes $p_\phi$ to
\begin{align}
p_\phi=-\frac{1}{2}\partial_z\Lambda.
\end{align}
Here, we could allow for some $y$-independent integration function, but such a function can always be set to zero by an appropriate redefinition of $\Lambda$. This integrability condition also tells us that 
\begin{align}
\partial_z\Omega_\mathcal{C}=ip_z \Omega_\mathcal{C}, \qquad \partial_y p_z=0.
\end{align}

At last, we consider the $d_6\bj$ and $d_6P$ equations given in \eqref{dJ} and \eqref{d6P}, respectively. The first one reduces to
\begin{align}
d_\mathcal{C} \left(e^{2w+2F -3\lambda} J_\mathcal{C} \right)&=e^{2w+2F-3\lambda}J_\mathcal{C}\wedge\partial_z V^R  \\
\partial_z\left(e^{4w} J_\mathcal{C} \wedge J_\mathcal{C} \right) &= 2 e^{2F} e^{2w -3\lambda} J_\mathcal{C} \wedge d_\mathcal{C} V^I\\
J_\mathcal{C}\wedge d_\mathcal{C} V^R&=0,
\end{align}
while the second one becomes
\begin{equation}
\frac{1}{y} \partial_y\left( e^{4w}  \right) J_\mathcal{C} \wedge J_\mathcal{C} = - e^{2w -3\lambda}  \left[d_\mathcal{C} P_\mathcal{C} - \partial_z \Lambda d_\mathcal{C} V^I \right]\wedge J_\mathcal{C} - \frac{1}{2} e^{4w -2F} \left(\frac{2}{y} \partial_y e^{2F} - \partial_z^2 \Lambda \right)  J_\mathcal{C} \wedge J_\mathcal{C}.
\end{equation}

%%%%%%%%%%%%%%%%%%%%%%%%%%%%%%%%%%
\subsection*{The canonical system}
%%%%%%%%%%%%%%%%%%%%%%%%%%%%%%%%%%%

Recall that we are looking at $AdS_3$ solutions where the internal manifold $M_8$ admits a $U(1)^2$ structure group. More particularly, recall that this condition led us to the conditions
\begin{equation}
p_z=0,\qquad V^R = d_{\mathcal{C}} \Gamma,
\end{equation} where $\Gamma$ is some generic function. To eliminate $V^R$ from the metric we make the coordinate transformation from $(\hat{x}^i, z, y)$ to $(x^i,u,t)$ defined by
\begin{equation}
x^i = \hat{x}^i, \qquad 2 t = y^2, \qquad z = -\Gamma(x^i,t,u).  
\end{equation} The differential forms and differential operators transform as
\begin{align}
 dx^i &= d\hat{x}^i , \qquad dt = y dy, \qquad \eta_z = - \partial_u \Gamma du - \partial_t \Gamma dt, \\
d_{\mathcal{C}} &= dx^i \partial_i, \qquad \partial_z = - \frac{1}{\partial_u \Gamma} \partial_u, \qquad y \partial_y = 2t \left(\partial_t - \frac{\partial_t \Gamma}{\partial_u \Gamma} \partial_u\right). 
\end{align}

%%%%%%%%%%%%%%%%%%%%%%%%%%%%%%%%%%%%%%%%%%%%%%%%%%
%\subsection{The metric for $U(1)^2$ systems}\label{U(1)sqmetric} The BPS equations can be reduced for the ansatz in equations \eqref{metans}-\eqref{omans}. In doing so, we find that the system is governed by three functions $(\Lambda, \Gamma, G)$ and two connection forms $(V^I, P_\mathcal{C})$ that live on the four-manifold $M_\mathcal{C}$.  All functions and forms depend on the $(t,u)$ and $x^i$, the coordinates on $M_\mathcal{C}$.  The functions $(\Lambda,G)$ are related to the ones above as   

%We can also add a term to $P$ that is proportional to $(\partial_u \Gamma du+\partial_t \Gamma dt)$.  This would imply that $\partial_u \Omega_4 = i \partial_u P_z \Omega_4$, for some function $P_z$, and that the killing spinor %is charged along a non-isometry direction, $\partial_u$.  We disallow such term since $u$ is generically not an isometry directions and since the $U(1)$'s are only fiberred over the surface $M_\mathcal{C}$. 

%%%%%%%%%%%%%%%%%%%%%%%%%%%%%%%%%%%%%%%%%%%%%%%%%%%

Now in these new coordinates, the metric reads as
\begin{align}
ds^2_{11} &= e^{2\lambda} \left[ ds^2_{AdS_3} + e^{2w -3\lambda} ds^2(M_\mathcal{C}) +\frac{1}{2} e^{-6\lambda} ds^2(N_4) \right] \\
ds^2(N_4) &= -\frac{G}{\partial_u \Gamma} \eta_\phi^2 - \frac{4 \partial_u \Gamma}{\det(h)} \left(d\psi +P \right)^2 - G\partial_u \Gamma \left(du +\frac{\partial_t \Gamma}{\partial_u \Gamma} dt \right)^2 - \frac{\det(g)}{\partial_u \Gamma} dt^2,
\end{align} 
\begin{align}
\mbox{where} \qquad \qquad  G&= - 2 e^{2F} \partial_u \Gamma,\qquad y\partial_y\Lambda= -\frac{2}{\sin^2 2\beta}, \\
\det(g) &= \partial_t \Lambda \partial_u \Gamma - \partial_t \Gamma \partial_u \Lambda \\
\det(h) &= \partial_t \left(\Lambda + \ln t\right)\partial_u \Gamma - \partial_t \Gamma \partial_u \Lambda.  
\end{align} The warp factors become
\begin{equation}
e^{-6\lambda} = \frac{1}{8t} \frac{\det(h)}{\det(g)}, \qquad e^{4 w} = \frac{t^2}{2G} \det(g) e^\Lambda,
\end{equation} and the connection form for the $S^1_\psi$ becomes
\begin{equation}
P =\frac{1}{2} P_\mathcal{C} +\frac{1}{2} \frac{\partial_u \Lambda}{\partial_u \Gamma} \eta_\phi .
\end{equation}
The $SU(2)$-structure, $(J_\mathcal{C},\Omega_\mathcal{C})$, of $M_\mathcal{C}$ satisfies
\begin{align}
d_\mathcal{C} J_{\mathcal{C}} = -d_\mathcal{C} \ln\left(G e^{2w-3\lambda}\right) \wedge J_\mathcal{C} \label{Jeqder}, \qquad d_\mathcal{C} \Omega_{\mathcal{C}} = \frac{1}{2}\left(i P_\mathcal{C} - d_\mathcal{C} \Lambda\right) \wedge \Omega_\mathcal{C}
\end{align} and 
\begin{equation}
\partial_t \left(J_\mathcal{C} \wedge J_\mathcal{C} \right)= \partial_u \left(J_\mathcal{C} \wedge J_\mathcal{C} \right)=0, \qquad \partial_t \Omega_\mathcal{C} = \partial_u \Omega_\mathcal{C}  =0.  
\end{equation} The connection forms satisfy the following holomorphicity conditions:
\begin{align}
\Omega_\mathcal{C} \wedge d_\mathcal{C} V^I &=\Omega_\mathcal{C} \wedge d_\mathcal{C} P_{\mathcal{C}} =0 \label{hol1}\\
\Omega_\mathcal{C} \wedge \partial_t \left( d_\mathcal{C} \Gamma+ i V^I\right)&= \Omega_\mathcal{C} \wedge  \partial_u \left( d_\mathcal{C}\Gamma+i V^I\right) = 0, \label{hol2} \\
\Omega_\mathcal{C} \wedge \partial_t \left(i P_\mathcal{C} - d_\mathcal{C} \Lambda\right)&=  \Omega_\mathcal{C} \wedge \partial_u \left(i P_\mathcal{C} - d_\mathcal{C} \Lambda\right) =0.  \label{hol3}
\end{align} We also have the following equations for the connection forms and the $e^{4w}$ warp factor:
\begin{align}
\partial_u e^{4w} J_\mathcal{C} \wedge J_\mathcal{C}  &= G e^{2w-3\lambda}  J_\mathcal{C} \wedge d_\mathcal{C} V^I  \label{uweq}\\ 
\partial_t e^{4w} J_\mathcal{C} \wedge J_\mathcal{C}  &= e^{2w-3\lambda}  J_\mathcal{C} \wedge \left[ G_2 d_\mathcal{C} V^I- d_\mathcal{C} P_\mathcal{C} \right] - \frac{e^{4w}}{G} \left[ \partial_t G - \partial_u  G_2 \right] J_\mathcal{C} \wedge J_\mathcal{C}   \label{tweq}
\end{align} where
\begin{equation}
G_2 = \frac{G \partial_t \Gamma -\partial_u \Lambda}{ \partial_u \Gamma}.  
\end{equation}
\newpage
At last, the magnetic flux is given by\footnote{Note the decomposition $\star_6=\star_4\star_2$ and furthermore, $\star_4 A_n= e^{2(2-n)w} \star_\mathcal{C} A_n$ where $A_n$ is an $n$-form on $M_\mathcal{C}$.}
\begin{equation}
\begin{split}
B_4=&B_4^{(1)}+B_4^{(2)}+B_4^{(3)}+B_4^{(4)}\\
B_4^{(1)}=&\frac{y}{2}e^{4w}\frac{(\partial_u G_2-\partial_t G)}{G}J_\mathcal{C}\wedge J_\mathcal{C}\\
&+\frac{y}{2}G \Bigg[e^{2w-3\lambda}G(\partial_u G_2-\partial_t G)J_\mathcal{C}+\frac{\partial_t\Gamma}{\partial_u\Gamma}\left(\partial_u(e^{2w-3\lambda}\jc)-\frac{1}{2}Ge^{-6\lambda}\dc V^I\right)\\
&-\left(\partial_t(e^{2w-3\lambda}\jc)+\frac{1}{2}e^{-6\lambda}\dc \pc\right)+\frac{1}{2}G_2e^{-6\lambda}\dc V^{I}\Bigg]\wedge\frac{\eta_z}{\partial_u\Gamma}\wedge\eta_\phi\\
&+\frac{y}{2}e^{2w-3\lambda}\left[G_2\partial_u V^I-G\partial_tV^I-\partial_u\pc\right]\wedge\jc\wedge\frac{\eta_z}{\partial_u\Gamma}\\
&+\frac{y}{2}e^{2w-3\lambda}\left[\dc G_2-\frac{\partial_t\Gamma}{\partial_u\Gamma}\dc G\right]\wedge\jc\wedge\eta_\phi\\
B_4^{(2)}=&\frac{e^{-6\lambda}}{\sin^22\beta}\Bigg\{e^{2w+3\lambda}\star_\mathcal{C}\dc\ln G\\
&+\frac{1}{\partial_u\Gamma}\left[e^{2w+3\lambda} \partial_u \ln e^{6\lambda} \,\jc+e^{6\lambda}\star_\mathcal{C}\left(\partial_u(e^{2w-3\lambda}\jc)-\frac{1}{2}Ge^{-6\lambda}\dc V^I\right)\right]\wedge\eta_\phi\\
&+\frac{1}{2}G\star_\mathcal{C}\left(\dc\ln(Ge^{-6\lambda})\wedge\jc\right)\wedge\frac{\eta_z}{\partial_u\Gamma}\wedge\eta_\phi\Bigg\}\wedge dy\\
B_4^{(3)}=&-y\Bigg\{e^{2w-3\lambda}\dc\ln G \wedge\jc+\left[\partial_u(e^{2w-3\lambda}\jc)-\frac{1}{2}Ge^{-6\lambda}\dc V^I\right]\wedge\frac{\eta_z}{\partial_u\Gamma}\\
&+\frac{1}{2}Ge^{-6\lambda}\dc\ln(Ge^{-6\lambda})\wedge\frac{\eta_z}{\partial_u\Gamma}\wedge\eta_\phi\Bigg\}\wedge D\psi\\
B_4^{(4)}=&\frac{1}{2}e^{-3\lambda}\Bigg\{G\left[\frac{1}{4}\frac{(\partial_u G_2-\partial_t G)}{G}-\frac{1}{4}\sin^22\beta \left(\partial_t-\frac{\partial_t\Gamma}{\partial_u\Gamma}\partial_u\right)\ln e^{4w}-e^{-3\lambda}\right]\wedge\frac{\eta_z}{\partial_u\Gamma}\wedge\eta_\phi\\
&-\frac{1}{4}\star_\mathcal{C}\left[G\sin^22\beta \left(\partial_tV^I-\frac{\partial_t\Gamma}{\partial_u\Gamma}\partial_uV^I\right)\wedge\jc-\Bigg(\partial_u P_\mathcal{C}+(G\partial_t-G_2\partial_u)V^I\Bigg)\wedge\jc\right]\wedge\frac{\eta_\phi}{\partial_u\Gamma}\\
&+\frac{1}{4}\star_\mathcal{C}\left[\left(\dc G_2-\frac{\partial_t\Gamma}{\partial_u\Gamma}\dc G\right)\wedge\jc-G\sin^22\beta\, \dc\left(\frac{\partial_t\Gamma}{\partial_u\Gamma}\right)\wedge\jc\right]\wedge\eta_z\\
&+2e^{2w+3\lambda}\left[e^{-3\lambda}+\frac{1}{8}\sin^22\beta \frac{2}{y}\partial_y\ln e^{2F}+\frac{1}{4}\frac{(\partial_u G_2-\partial_t G)}{G}\right]\jc\\
&+\frac{1}{2}e^{6\lambda}\star_\mathcal{C}\Bigg[\frac{\partial_t\Gamma}{\partial_u\Gamma}\left(\cos^22\beta\partial_u(e^{2w-3\lambda}\jc)-\frac{1}{2}Ge^{-6\lambda}\dc V^I\right)\\
&-\left(\cos^22\beta\partial_t(e^{2w-3\lambda}\jc)+\frac{1}{2}e^{-6\lambda}\dc \pc\right)+\frac{1}{2}G_2e^{-6\lambda}\dc V^{I}\Bigg]\Bigg\}\wedge dy\wedge D\psi\\
\end{split}
\end{equation}

In conclusion, the base four-manifold, $M_\mathcal{C}$ is conformally K\"ahler.  The metric on the $N_4$ fiber is determined by the functions $(G,\Lambda, \Gamma)$ while the forms $(P_{\mathcal{C}}, V^I)$ fix the connection of the bundle.  This system of equations can be further reduced by imposing an appropriate ansatz for the four-form flux, $B_4$.

%%%%%%%%%%%%%%%%%%%%%%%%%%%%%%%%%%%%%%%%%%%%%%
\section{Ansatz for the magnetic flux}\label{fluxapp}
%%%%%%%%%%%%%%%%%%%%%%%%%%%%%%%%%%%%%%%%%%%%%%%%
Recall that the different terms of $B_4$ can be organized in such a way that $B_4$ takes the form given in equation \eqref{B4formal}. Recall also our argument for setting $\mathcal{B}^{(4)}$ and $\mathcal{B}^{(3)}_a$ to zero. In this appendix we discuss the consequences of that. 

The vanishing of $\mathcal{B}^{(4)}$, i.e. the $\jc\wedge\jc$ term of $B_4^{(1)}$, implies that
 \begin{equation}
 \partial_t G = \partial_u G_2.  \label{intG}
 \end{equation} The vanishing of $\mathcal{B}^{(3)}_a$ implies that
 \begin{equation}
 d_\mathcal{C} G = d_\mathcal{C} G_2 =0, \qquad \partial_u P_\mathcal{C} +\left( G \partial_t - G_2 \partial_u \right) V^I =0.   \label{dG}
 \end{equation} The constraint on the one-forms $(P_\mathcal{C},V^I)$ also follows from the $(G,G_2)$ constraints and the holomorphicity constraints in \eqref{hol2} and \eqref{hol3}.  Since the $G$'s are independent of the base and satisfy \eqref{intG}, one can make a coordinate transformation\footnote{This coordinate transformation is defined by
 \begin{align}
 dt'=dt,\qquad du'=G_2dt+Gdu.
 \end{align}
 }  that removes them from the metric and the BPS equations. This is equivalent to fixing $(G=1,G_2=0)$.  The potentials and one-forms sastisfy
 \begin{equation}
 \partial_t \Gamma = \partial_u \Lambda, \qquad  \partial_t V^I = - \partial_u P_\mathcal{C}.  
 \end{equation} These can be solved in terms of a single potential $D_0$ and one-form $P_0$ which respect to which we have
 \begin{align}
  \Lambda &= \partial_t D_0, \qquad P_\mathcal{C}= 2\partial_t P_0 \\ 
 \Gamma &= \partial_u D_0, \qquad V^I =- 2\partial_u P_0.  
 \end{align} 
Notice that after imposing these constraints, the number of undetermined forms has been reduced from five down to two; namely, the zero-form $D_0$ and the one-form $P_0$. At last, the magnetic flux reduces to\newpage
\begin{equation}
\begin{split}
B_4=&B_4^{(1)}+B_4^{(2)}+B_4^{(3)}+B_4^{(4)} \label{fluxform} \\
B_4^{(1)}=&+\frac{y}{2} \Big(A_u-A_t\Big)\wedge\frac{\eta_z}{D_{uu}}\wedge\eta_\phi\\
B_4^{(2)}=&\frac{1}{\sin^22\beta}\Bigg\{\frac{1}{D_{uu}}\left[-e^{2w+3\lambda} \partial_u  (e^{-6\lambda}) \,\jc+\star_\mathcal{C}\frac{D_{uu}}{D_{tu}}A_u\right]\wedge\eta_\phi+\frac{1}{2}\star_\mathcal{C}\left(\dc(e^{-6\lambda})\wedge\jc\right)\wedge\frac{\eta_z}{D_{uu}}\wedge\eta_\phi\Bigg\}\wedge dy\\
B_4^{(3)}=&-y\Bigg\{\frac{D_{uu}}{D_{tu}}A_u\wedge\frac{\eta_z}{D_{uu}}+\frac{1}{2}\dc(e^{-6\lambda})\wedge\frac{\eta_z}{D_{uu}}\wedge\eta_\phi\Bigg\}\wedge D\psi\\
B_4^{(4)}=&\frac{1}{2}e^{-3\lambda}\Bigg\{-\left[\frac{1}{4}\sin^22\beta \left(\partial_t-\frac{D_{tu}}{D_{uu}}\partial_u\right)\ln e^{4w}+e^{-3\lambda}\right]\frac{\eta_z}{D_{uu}}\wedge\eta_\phi\\
&+\frac{1}{2}\star_\mathcal{C}\left[\sin^22\beta \left(\partial_t-\frac{D_{tu}}{D_{uu}}\partial_u\right)\partial_u P_0\wedge\jc\right]\wedge\frac{\eta_\phi}{D_{uu}}\\
&+\frac{1}{4}\star_\mathcal{C}\left[-\sin^22\beta\, \dc\left(\frac{D_{tu}}{D_{uu}}\right)\wedge\jc\right]\wedge\eta_z\\
&+2e^{2w+3\lambda}\left[e^{-3\lambda}+\frac{1}{8}\sin^22\beta \frac{2}{y}\partial_y\ln e^{2F}\right]\jc\\
&+\frac{1}{2}e^{6\lambda}\star_\mathcal{C}\Bigg[A_u-A_t+\sin^22\beta\left(\partial_t - \frac{D_{tu}}{D_{uu}}\partial_u\right)(e^{2w-3\lambda}\jc)\Bigg]\Bigg\}\wedge dy\wedge D\psi,
\end{split}
\end{equation}
where $D_{tu}=\partial_t\partial_u D_0$, etc, and
\begin{align}
A_u=&\frac{D_{tu}}{D_{uu}}\left[\partial_u(e^{2w-3\lambda}\jc)+e^{-6\lambda}\partial_u(\dc P_0)\right]\\
A_t=&\partial_t(e^{2w-3\lambda}\jc)+e^{-6\lambda}\partial_t(\dc P_0).
\end{align}

%%%%%%%%%%%%%%%%%%%%%%%%%%%%%%%%%%%%%
\subsection*{The $\pm$ coordinates}
%%%%%%%%%%%%%%%%%%%%%%%%%%%%%%%%%%%%%%%

 As discussed in chapter 3, there is a natural coordinate system on $N_4$ where the circles $S_\pm ^1$ become apparent. Here we present in detail how the system can be put in this form.

After imposing the flux constraints, the metric on the $N_4$ fiber becomes
 \begin{align}
 ds^2(N_4)=&-(D_{tt}dt^2+2D_{tu}dtdu+D_{uu}du^2)\\
 &-\frac{4}{\det(h)}\left[D_{uu}\eta_t^2-2D_{tu}\eta_t\eta_u+\partial_t(D_{t}+\ln t)\eta_u^2\right].
 \end{align}
 Here we have introduced the one-forms $\eta_t$ and $\eta_u$; these are the natural objects appearing in the metric and they are defined as
 \begin{align}
\eta_u=-\frac{1}{2}\eta_\phi,\qquad \eta_t=D\psi-\frac{1}{2}\frac{D_{tu}}{D_{uu}}\eta_\phi,
 \end{align}
 or, in terms of $(D_0,P_0)$, as
  \begin{align}
\eta_u=-\frac{1}{2}d\phi+\partial_uP_0,\qquad \eta_t=d\psi+\partial_tP_0. 
 \end{align}
 The functions $\det(g)$ and $\det(h)$ become
 \begin{align}
 \det(g)=D_{tt}D_{uu}-(D_{tu})^2,\qquad \det(h)=\partial_t(D_{t}+\ln t)D_{uu}-(D_{tu})^2.
 \end{align}
 Next, we notice that this metric can be written in a more compact form:
 \begin{align}
 ds^2(N_4)=&g_{ij}dy^idy^j+4h^{ij}\eta_i\eta_j\\
 g_{ij}=&-\partial_i\partial_j D_0\\
 h_{ij}=&-\partial_i\partial_j [D_0+t(\ln t-1)],
 \end{align}
 with $i,j\in\{t,u\}$, and $y^t=t,y^u=u$. The matrix $h^{ij}$ is the inverse of $h_{ij}$. Moreover, the so far abstract notation of the functions $\det(g)$ and $\det(h)$ now acquires precise meaning, i.e. these functions denote the determinant of the metrics $g_{ij}$ and $h_{ij}$, respectively.
 
Now, this set up allows us to consider the following coordinate transformation:
\begin{align}
\psi=&\phi_++\phi_-,\qquad \frac{1}{2}\phi=a_+\phi_--a_-\phi_+,\\
u=&t^+-t^-,\quad\,\,\, \qquad t=a_+t^++a_-t^-,
\end{align}
 with $a_++a_-=1$. The differential operators and the $\eta$ one-forms transform as
 \begin{align}
\partial_\phi=&-\frac{1}{2}(\partial_{\phi_+}-\partial_{\phi_-}),\qquad  \partial_\psi=a_+\partial_{\phi_+}+a_-\partial_{\phi_-},\\
 \partial_t=&\partial_++\partial_-,\qquad \partial_u=a_-\partial_+-a_+\partial_-\\
  \eta_t=&\eta_++\eta_-,\qquad \eta_u=a_-\eta_+-a_+\eta_-,
 \end{align}
 where
 \begin{align}
 \eta_\pm=d\phi_\pm+\partial_{\pm} P_0.
 \end{align}
 At last, the full metric expressed in these coordinates is described in subsection \ref{02system}.

\bibliographystyle{utphys}
\bibliography{miracle}
\end{document}